\documentclass[longauth]{aa}  

\usepackage{graphicx}
\usepackage{txfonts}
\usepackage[pdfstartview=FitH,      
            breaklinks=true,        
            bookmarksopen=true,     
            bookmarksnumbered=true,  
            colorlinks=true,
            linkcolor=black,
            urlcolor=blue,
            citecolor=blue]{hyperref}
\usepackage{capt-of}
\usepackage{color}

\begin{document}

   \title{Examining the orbital decay targets KELT-9\,b, KELT-16\,b and WASP-4\,b, and the transit-timing variations of HD\,97658\,b\thanks{This article uses data from CHEOPS programme CH\_PR100012 and CH\_PR100013. Photometry data according to Table~\ref{tab:obs_log}, as well as the full Tables~\ref{tab:KELT-9_tts}-\ref{tab:WASP-4_tts} and Table~\ref{tab:HD97658_tts} are available at the CDS via anonymous ftp to cdsarc.u-strasbg.fr (130.79.128.5) or via \href{http://cdsarc.u-strasbg.fr/viz-bin/qcat?J/A+A/}{http://cdsarc.u-strasbg.fr/viz-bin/qcat?J/A+A/}}}

   \author{J.-V. Harre\inst{1}\thanks{E-mail: jan-vincent.harre@dlr.de} 
   $^{\href{https://orcid.org/0000-0001-8935-2472}{\includegraphics[scale=0.5]{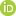}}}$ \and 
A. M. S. Smith\inst{1} $^{\href{https://orcid.org/0000-0002-2386-4341}{\includegraphics[scale=0.5]{figures/orcid.jpg}}}$ \and
S. C. C. Barros\inst{2,3} $^{\href{https://orcid.org/0000-0003-2434-3625}{\includegraphics[scale=0.5]{figures/orcid.jpg}}}$ \and
G. Boué\inst{4} $^{\href{https://orcid.org/0000-0002-5057-7743}{\includegraphics[scale=0.5]{figures/orcid.jpg}}}$ \and
Sz. Csizmadia\inst{1} $^{\href{https://orcid.org/0000-0001-6803-9698}{\includegraphics[scale=0.5]{figures/orcid.jpg}}}$ \and
D. Ehrenreich\inst{5} $^{\href{https://orcid.org/https://www.notion.so/D-Ehrenreich-7e5007d8e19f46368c036b8dfbbd5a59}{\includegraphics[scale=0.5]{figures/orcid.jpg}}}$ \and
H.-G. Florén\inst{6} \and 
A. Fortier\inst{7,8} $^{\href{https://orcid.org/0000-0001-8450-3374}{\includegraphics[scale=0.5]{figures/orcid.jpg}}}$ \and
P. F. L. Maxted\inst{9} $^{\href{https://orcid.org/0000-0003-3794-1317}{\includegraphics[scale=0.5]{figures/orcid.jpg}}}$ \and
M. J. Hooton\inst{10,11} $^{\href{https://orcid.org/0000-0003-0030-332X}{\includegraphics[scale=0.5]{figures/orcid.jpg}}}$ \and
B. Akinsanmi\inst{2,5} $^{\href{https://orcid.org/0000-0001-6519-1598}{\includegraphics[scale=0.5]{figures/orcid.jpg}}}$ \and
L. M. Serrano\inst{12} $^{\href{https://orcid.org/0000-0001-9211-3691}{\includegraphics[scale=0.5]{figures/orcid.jpg}}}$ \and
N. M. Rosário\inst{2,3} \and 
B.-O. Demory\inst{8} $^{\href{https://orcid.org/0000-0002-9355-5165}{\includegraphics[scale=0.5]{figures/orcid.jpg}}}$ \and
K. Jones\inst{8} \and 
J. Laskar\inst{4} $^{\href{https://orcid.org/0000-0003-2634-789X}{\includegraphics[scale=0.5]{figures/orcid.jpg}}}$ \and
V. Adibekyan\inst{2} $^{\href{https://orcid.org/0000-0002-0601-6199}{\includegraphics[scale=0.5]{figures/orcid.jpg}}}$ \and
Y. Alibert\inst{7} $^{\href{https://orcid.org/0000-0002-4644-8818}{\includegraphics[scale=0.5]{figures/orcid.jpg}}}$ \and
R. Alonso\inst{13,14} $^{\href{https://orcid.org/0000-0001-8462-8126}{\includegraphics[scale=0.5]{figures/orcid.jpg}}}$ \and
D. R. Anderson\inst{15} \and 
G. Anglada\inst{16,17} $^{\href{https://orcid.org/0000-0002-3645-5977}{\includegraphics[scale=0.5]{figures/orcid.jpg}}}$ \and
J. Asquier\inst{18} \and 
T. Bárczy\inst{19} $^{\href{https://orcid.org/0000-0002-7822-4413}{\includegraphics[scale=0.5]{figures/orcid.jpg}}}$ \and
D. Barrado y Navascues\inst{20} $^{\href{https://orcid.org/0000-0002-5971-9242}{\includegraphics[scale=0.5]{figures/orcid.jpg}}}$ \and
W. Baumjohann\inst{21} $^{\href{https://orcid.org/0000-0001-6271-0110}{\includegraphics[scale=0.5]{figures/orcid.jpg}}}$ \and
M. Beck\inst{5} $^{\href{https://orcid.org/0000-0003-3926-0275}{\includegraphics[scale=0.5]{figures/orcid.jpg}}}$ \and
T. Beck\inst{7} \and 
W. Benz\inst{7,8} $^{\href{https://orcid.org/0000-0001-7896-6479}{\includegraphics[scale=0.5]{figures/orcid.jpg}}}$ \and
N. Billot\inst{5} $^{\href{https://orcid.org/0000-0003-3429-3836}{\includegraphics[scale=0.5]{figures/orcid.jpg}}}$ \and
F. Biondi\inst{22,23} $^{\href{https://orcid.org/0000-0002-1337-3653}{\includegraphics[scale=0.5]{figures/orcid.jpg}}}$ \and
A. Bonfanti\inst{21} $^{\href{https://orcid.org/0000-0002-1916-5935}{\includegraphics[scale=0.5]{figures/orcid.jpg}}}$ \and
X. Bonfils\inst{24} $^{\href{https://orcid.org/0000-0001-9003-8894}{\includegraphics[scale=0.5]{figures/orcid.jpg}}}$ \and
A. Brandeker\inst{6} $^{\href{https://orcid.org/0000-0002-7201-7536}{\includegraphics[scale=0.5]{figures/orcid.jpg}}}$ \and
C. Broeg\inst{7,25} $^{\href{https://orcid.org/0000-0001-5132-2614}{\includegraphics[scale=0.5]{figures/orcid.jpg}}}$ \and
J. Cabrera\inst{1} \and 
V. Cessa\inst{7} \and 
S. Charnoz\inst{26} $^{\href{https://orcid.org/0000-0002-7442-491X}{\includegraphics[scale=0.5]{figures/orcid.jpg}}}$ \and
A. Collier Cameron\inst{27} $^{\href{https://orcid.org/0000-0002-8863-7828}{\includegraphics[scale=0.5]{figures/orcid.jpg}}}$ \and
M. B. Davies\inst{28} $^{\href{https://orcid.org/0000-0001-6080-1190}{\includegraphics[scale=0.5]{figures/orcid.jpg}}}$ \and
M. Deleuil\inst{29} $^{\href{https://orcid.org/0000-0001-6036-0225}{\includegraphics[scale=0.5]{figures/orcid.jpg}}}$ \and
L. Delrez\inst{30,31} $^{\href{https://orcid.org/0000-0001-6108-4808}{\includegraphics[scale=0.5]{figures/orcid.jpg}}}$ \and
O. D. S. Demangeon\inst{2,3} $^{\href{https://orcid.org/0000-0001-7918-0355}{\includegraphics[scale=0.5]{figures/orcid.jpg}}}$ \and
A. Erikson\inst{1} \and 
L. Fossati\inst{21} $^{\href{https://orcid.org/0000-0003-4426-9530}{\includegraphics[scale=0.5]{figures/orcid.jpg}}}$ \and
M. Fridlund\inst{32,33} $^{\href{https://orcid.org/0000-0002-0855-8426}{\includegraphics[scale=0.5]{figures/orcid.jpg}}}$ \and
D. Gandolfi\inst{12} $^{\href{https://orcid.org/0000-0001-8627-9628}{\includegraphics[scale=0.5]{figures/orcid.jpg}}}$ \and
M. Gillon\inst{30} $^{\href{https://orcid.org/0000-0003-1462-7739}{\includegraphics[scale=0.5]{figures/orcid.jpg}}}$ \and
M. Güdel\inst{34} \and 
C. Hellier\inst{9} $^{\href{https://orcid.org/0000-0002-3439-1439}{\includegraphics[scale=0.5]{figures/orcid.jpg}}}$ \and
K. Heng\inst{8,35} $^{\href{https://orcid.org/0000-0003-1907-5910}{\includegraphics[scale=0.5]{figures/orcid.jpg}}}$ \and
S. Hoyer\inst{29} $^{\href{https://orcid.org/0000-0003-3477-2466}{\includegraphics[scale=0.5]{figures/orcid.jpg}}}$ \and
K. G. Isaak\inst{36} $^{\href{https://orcid.org/0000-0001-8585-1717}{\includegraphics[scale=0.5]{figures/orcid.jpg}}}$ \and
L. L. Kiss\inst{37,38} \and 
A. Lecavelier des Etangs\inst{39} $^{\href{https://orcid.org/0000-0002-5637-5253}{\includegraphics[scale=0.5]{figures/orcid.jpg}}}$ \and
M. Lendl\inst{5} $^{\href{https://orcid.org/0000-0001-9699-1459}{\includegraphics[scale=0.5]{figures/orcid.jpg}}}$ \and
C. Lovis\inst{5} $^{\href{https://orcid.org/0000-0001-7120-5837}{\includegraphics[scale=0.5]{figures/orcid.jpg}}}$ \and
A. Luntzer\inst{40} \and 
D. Magrin\inst{22} $^{\href{https://orcid.org/0000-0003-0312-313X}{\includegraphics[scale=0.5]{figures/orcid.jpg}}}$ \and
V. Nascimbeni\inst{22} $^{\href{https://orcid.org/0000-0001-9770-1214}{\includegraphics[scale=0.5]{figures/orcid.jpg}}}$ \and
G. Olofsson\inst{6} $^{\href{https://orcid.org/0000-0003-3747-7120}{\includegraphics[scale=0.5]{figures/orcid.jpg}}}$ \and
R. Ottensamer\inst{40} \and 
I. Pagano\inst{41} $^{\href{https://orcid.org/0000-0001-9573-4928}{\includegraphics[scale=0.5]{figures/orcid.jpg}}}$ \and
E. Pallé\inst{13} $^{\href{https://orcid.org/0000-0003-0987-1593}{\includegraphics[scale=0.5]{figures/orcid.jpg}}}$ \and
C. M. Persson\inst{33} \and 
G. Peter\inst{42} $^{\href{https://orcid.org/0000-0001-6101-2513}{\includegraphics[scale=0.5]{figures/orcid.jpg}}}$ \and
G. Piotto\inst{22,43} $^{\href{https://orcid.org/0000-0002-9937-6387}{\includegraphics[scale=0.5]{figures/orcid.jpg}}}$ \and
D. Pollacco\inst{35} \and 
D. Queloz\inst{44,10} $^{\href{https://orcid.org/0000-0002-3012-0316}{\includegraphics[scale=0.5]{figures/orcid.jpg}}}$ \and
R. Ragazzoni\inst{22,43} $^{\href{https://orcid.org/0000-0002-7697-5555}{\includegraphics[scale=0.5]{figures/orcid.jpg}}}$ \and
N. Rando\inst{18} \and 
H. Rauer\inst{1,45,46} $^{\href{https://orcid.org/0000-0002-6510-1828}{\includegraphics[scale=0.5]{figures/orcid.jpg}}}$ \and
I. Ribas\inst{16,17} $^{\href{https://orcid.org/0000-0002-6689-0312}{\includegraphics[scale=0.5]{figures/orcid.jpg}}}$ \and
G. R. Ricker\inst{47,48} $^{\href{https://orcid.org/0000-0003-2058-6662}{\includegraphics[scale=0.5]{figures/orcid.jpg}}}$ \and
S. Salmon\inst{5} $^{\href{https://orcid.org/0000-0002-1714-3513}{\includegraphics[scale=0.5]{figures/orcid.jpg}}}$ \and
N. C. Santos\inst{2,3} $^{\href{https://orcid.org/0000-0003-4422-2919}{\includegraphics[scale=0.5]{figures/orcid.jpg}}}$ \and
G. Scandariato\inst{41} $^{\href{https://orcid.org/0000-0003-2029-0626}{\includegraphics[scale=0.5]{figures/orcid.jpg}}}$ \and
S. Seager\inst{49,50,51} $^{\href{https://orcid.org/0000-0002-6892-6948}{\includegraphics[scale=0.5]{figures/orcid.jpg}}}$ \and
D. Ségransan\inst{5} $^{\href{https://orcid.org/0000-0003-2355-8034}{\includegraphics[scale=0.5]{figures/orcid.jpg}}}$ \and
A. E. Simon\inst{7} $^{\href{https://orcid.org/0000-0001-9773-2600}{\includegraphics[scale=0.5]{figures/orcid.jpg}}}$ \and
S. G. Sousa\inst{2} $^{\href{https://orcid.org/0000-0001-9047-2965}{\includegraphics[scale=0.5]{figures/orcid.jpg}}}$ \and
M. Steller\inst{21} $^{\href{https://orcid.org/0000-0003-2459-6155}{\includegraphics[scale=0.5]{figures/orcid.jpg}}}$ \and
Gy. M. Szabó\inst{52,53} \and 
N. Thomas\inst{7} \and 
S. Udry\inst{5} $^{\href{https://orcid.org/0000-0001-7576-6236}{\includegraphics[scale=0.5]{figures/orcid.jpg}}}$ \and
B. Ulmer\inst{42} \and 
V. Van Grootel\inst{31} $^{\href{https://orcid.org/0000-0003-2144-4316}{\includegraphics[scale=0.5]{figures/orcid.jpg}}}$ \and
N. A. Walton\inst{54} $^{\href{https://orcid.org/0000-0003-3983-8778}{\includegraphics[scale=0.5]{figures/orcid.jpg}}}$ \and
T. G. Wilson\inst{27} $^{\href{https://orcid.org/0000-0001-8749-1962}{\includegraphics[scale=0.5]{figures/orcid.jpg}}}$ \and
J. N. Winn\inst{55} $^{\href{https://orcid.org/0000-0002-4265-047X}{\includegraphics[scale=0.5]{figures/orcid.jpg}}}$ \and
B. Wohler\inst{56,57} $^{\href{https://orcid.org/0000-0002-5402-9613}{\includegraphics[scale=0.5]{figures/orcid.jpg}}}$
}

\institute{\label{inst:1} Institute of Planetary Research, German Aerospace Center (DLR), Rutherfordstrasse 2, 12489 Berlin, Germany \and
\label{inst:2} Instituto de Astrofisica e Ciencias do Espaco, Universidade do Porto, CAUP, Rua das Estrelas, 4150-762 Porto, Portugal \and
\label{inst:3} Departamento de Fisica e Astronomia, Faculdade de Ciencias, Universidade do Porto, Rua do Campo Alegre, 4169-007 Porto, Portugal \and
\label{inst:4} IMCCE, UMR8028 CNRS, Observatoire de Paris, PSL Univ., Sorbonne Univ., 77 av. Denfert-Rochereau, 75014 Paris, France \and
\label{inst:5} Observatoire Astronomique de l'Université de Genève, Chemin Pegasi 51, CH-1290 Versoix, Switzerland \and
\label{inst:6} Department of Astronomy, Stockholm University, AlbaNova University Center, 10691 Stockholm, Sweden \and
\label{inst:7} Physikalisches Institut, University of Bern, Sidlerstrasse 5, 3012 Bern, Switzerland \and
\label{inst:8} Center for Space and Habitability, University of Bern, Gesellschaftsstrasse 6, 3012 Bern, Switzerland \and
\label{inst:9} Astrophysics Group, Keele University, Staffordshire, ST5 5BG, United Kingdom \and
\label{inst:10} Cavendish Laboratory, JJ Thomson Avenue, Cambridge CB3 0HE, UK \and
\label{inst:11} Physikalisches Institut, University of Bern, Gesellsschaftstrasse 6, 3012 Bern, Switzerland \and
\label{inst:12} Dipartimento di Fisica, Universita degli Studi di Torino, via Pietro Giuria 1, I-10125, Torino, Italy \and
\label{inst:13} Instituto de Astrofisica de Canarias, 38200 La Laguna, Tenerife, Spain \and
\label{inst:14} Departamento de Astrofisica, Universidad de La Laguna, 38206 La Laguna, Tenerife, Spain \and
\label{inst:15} Department of Physics, University of Warwick, Gibbet Hill Road, Coventry, CV4 7AL, United Kingdom \and
\label{inst:16} Institut de Ciencies de l'Espai (ICE, CSIC), Campus UAB, Can Magrans s/n, 08193 Bellaterra, Spain \and
\label{inst:17} Institut d'Estudis Espacials de Catalunya (IEEC), 08034 Barcelona, Spain \and
\label{inst:18} ESTEC, European Space Agency, 2201AZ, Noordwijk, NL \and
\label{inst:19} Admatis, 5. Kandó Kálmán Street, 3534 Miskolc, Hungary \and
\label{inst:20} Depto. de Astrofisica, Centro de Astrobiologia (CSIC-INTA), ESAC campus, 28692 Villanueva de la Cañada (Madrid), Spain \and
\label{inst:21} Space Research Institute, Austrian Academy of Sciences, Schmiedlstrasse 6, A-8042 Graz, Austria \and
\label{inst:22} INAF, Osservatorio Astronomico di Padova, Vicolo dell'Osservatorio 5, 35122 Padova, Italy \and
\label{inst:23} Max-Planck-Institut für Extraterrestrische Physik, Giessenbachstr. 1, D-85748 Garching, Germany \and
\label{inst:24} Université Grenoble Alpes, CNRS, IPAG, 38000 Grenoble, France \and
\label{inst:25} Center for Space and Habitability, Gesellsschaftstrasse 6, 3012 Bern, Switzerland \and
\label{inst:26} Université de Paris, Institut de physique du globe de Paris, CNRS, F-75005 Paris, France \and
\label{inst:27} Centre for Exoplanet Science, SUPA School of Physics and Astronomy, University of St Andrews, North Haugh, St Andrews KY16 9SS, UK \and
\label{inst:28} Centre for Mathematical Sciences, Lund University, Box 118, 221 00 Lund, Sweden \and
\label{inst:29} Aix Marseille Univ, CNRS, CNES, LAM, 38 rue Frédéric Joliot-Curie, 13388 Marseille, France \and
\label{inst:30} Astrobiology Research Unit, Université de Liège, Allée du 6 Août 19C, B-4000 Liège, Belgium \and
\label{inst:31} Space sciences, Technologies and Astrophysics Research (STAR) Institute, Université de Liège, Allée du 6 Août 19C, 4000 Liège, Belgium \and
\label{inst:32} Leiden Observatory, University of Leiden, PO Box 9513, 2300 RA Leiden, The Netherlands \and
\label{inst:33} Department of Space, Earth and Environment, Chalmers University of Technology, Onsala Space Observatory, 439 92 Onsala, Sweden \and
\label{inst:34} University of Vienna, Department of Astrophysics, Türkenschanzstrasse 17, 1180 Vienna, Austria \and
\label{inst:35} Department of Physics, University of Warwick, Gibbet Hill Road, Coventry CV4 7AL, United Kingdom \and
\label{inst:36} Science and Operations Department - Science Division (SCI-SC), Directorate of Science, European Space Agency (ESA), European Space Research and Technology Centre (ESTEC),
Keplerlaan 1, 2201-AZ Noordwijk, The Netherlands \and
\label{inst:37} Konkoly Observatory, Research Centre for Astronomy and Earth Sciences, 1121 Budapest, Konkoly Thege Miklós út 15-17, Hungary \and
\label{inst:38} ELTE E\"otv\"os Lor\'and University, Institute of Physics, P\'azm\'any P\'eter s\'et\'any 1/A, 1117 \and
\label{inst:39} Institut d'astrophysique de Paris, UMR7095 CNRS, Université Pierre \& Marie Curie, 98bis blvd. Arago, 75014 Paris, France \and
\label{inst:40} Department of Astrophysics, University of Vienna, Tuerkenschanzstrasse 17, 1180 Vienna, Austria \and
\label{inst:41} INAF, Osservatorio Astrofisico di Catania, Via S. Sofia 78, 95123 Catania, Italy \and
\label{inst:42} Institute of Optical Sensor Systems, German Aerospace Center (DLR), Rutherfordstrasse 2, 12489 Berlin, Germany \and
\label{inst:43} Dipartimento di Fisica e Astronomia "Galileo Galilei", Universita degli Studi di Padova, Vicolo dell'Osservatorio 3, 35122 Padova, Italy \and
\label{inst:44} ETH Zurich, Department of Physics, Wolfgang-Pauli-Strasse 2, CH-8093 Zurich, Switzerland \and
\label{inst:45} Zentrum für Astronomie und Astrophysik, Technische Universität Berlin, Hardenbergstr. 36, D-10623 Berlin, Germany \and
\label{inst:46} Institut für Geologische Wissenschaften, Freie Universität Berlin, 12249 Berlin, Germany \and
\label{inst:47} MIT Kavli Institute for Astrophysics and Space Research, 70 Vassar St, Cambridge, MA 02139, USA \and
\label{inst:48} MIT Physics Department, 182 Memorial Dr, Cambridge, MA 02142, USA \and
\label{inst:49} Department of Earth, Atmospheric, and Planetary Sciences, Massachusetts Institute of Technology, Cambridge, MA 02139, USA \and
\label{inst:50} Department of Physics and Kavli Institute for Astrophysics and Space Research, Massachusetts Institute of Technology, Cambridge, MA 02139, USA \and
\label{inst:51} Department of Aeronautics and Astronautics, Massachusetts Institute of Technology, Cambridge, MA 02139, USA \and
\label{inst:52} ELTE E\"otv\"os Lor\'and University, Gothard Astrophysical Observatory, 9700 Szombathely, Szent Imre h. u. 112, Hungary \and
\label{inst:53} MTA-ELTE Exoplanet Research Group, 9700 Szombathely, Szent Imre h. u. 112, Hungary \and
\label{inst:54} Institute of Astronomy, University of Cambridge, Madingley Road, Cambridge, CB3 0HA, United Kingdom \and
\label{inst:55} Department of Astrophysical Sciences, Princeton University, Princeton, NJ 08544, USA \and
\label{inst:56} SETI Institute, Mountain View, CA  94043, USA \and
\label{inst:57} NASA Ames Research Center, Moffett Field, CA  94035, USA
}
\authorrunning{J.-V. Harre et al.}
\titlerunning{Examining the orbital decay targets KELT-9\,b, KELT-16\,b and WASP-4\,b, and the TTVs of HD\,97658\,b}
   \date{Received 07 18, 2022; accepted MM dd, 2022}

 
  \abstract
   {Tidal orbital decay is suspected to occur especially for hot Jupiters, with the only observationally confirmed case of this being WASP-12\,b. By examining this effect, information on the properties of the host star can be obtained using the so-called stellar modified tidal quality factor $Q_*'$, which describes the efficiency with which kinetic energy of the planet is dissipated within the star. This can help to get information about the interior of the star.}
   {In this study, we aim to improve constraints on the tidal decay of the KELT-9, KELT-16 and WASP-4 systems, to find evidence for or against the presence of this particular effect. With this, we want to constrain each star's respective $Q_*'$ value.
   In addition to that, we also aim to test the existence of the transit timing variations (TTVs) in the HD\,97658 system, which previously favoured a quadratic trend with increasing orbital period.}
   {Making use of newly acquired photometric observations from CHEOPS (CHaracterising ExOplanet Satellite) and TESS (Transiting Exoplanet Survey Satellite), combined with archival transit and occultation data, we use Markov chain Monte Carlo (MCMC) algorithms to fit three models, a constant period model, an orbital decay model, and an apsidal precession model, to the data.}
   {We find that the KELT-9 system is best described by an apsidal precession model for now, with an orbital decay trend at over $2\,\sigma$ being a possible solution as well. A Keplerian orbit model with a constant orbital period fits the transit timings of KELT-16\,b the best due to the scatter and scale of their error bars. The WASP-4 system is represented the best by an orbital decay model at a $5\,\sigma$ significance, although apsidal precession cannot be ruled out with the present data. 
   For HD\,97658\,b, using recently acquired transit observations, we find no conclusive evidence for a previously suspected strong quadratic trend in the data.
   }
   {}

   \keywords{Planets and satellites: dynamical evolution and stability --
                Planet-star interactions --
                Techniques: photometric
               }

   \maketitle
%

\section{Introduction}

Hot Jupiters, since they are so close to their host stars, are expected to tidally interact with them. This interaction is expressed in the planet raising a tidal bulge on the surface of the star, due to the gravitational attraction from the planet's mass. Vice-versa, the star also raises a bulge on the surface of the planet, which however, has a negligible influence on the effect we examine here, that is, tidal decay of the orbit of hot Jupiters. If the stellar rotation is not synchronized with the planetary orbital period, the viscosity of the star's plasma leads to a lag between the tide raised on the surface of the star and the planet-to-star center, which leads to a transfer of orbital angular momentum from the planet to the star, if the orbital period of the planet is smaller than the stellar rotation period. This is described as the equilibrium tide.
This subsequently leads to the planet to slowly spiral inwards and the star spinning up \citep{1973ApJ...180..307C,1996ApJ...470.1187R}.
An additional contribution is made from the dynamical tide, which arises from stellar oscillations \citep{Ogilvie_2014}.\\
The hitherto only planet for which tidal orbital decay is observationally confirmed, is WASP-12\,b. Discovered in the Wide Angle Search for Planets (WASP) project \citep{2006PASP..118.1407P} by \citet{2009ApJ...693.1920H}, it orbits its host star, a late F-type star, every 1.09 days, has a mass of 1.47 M$_\mathrm{J}$ and a radius of 1.90 R$_\mathrm{J}$. The variations in transit and occultation timings observed for this planet lead to a decrease of 29\,ms\,yr$^{-1}$ in its orbital period, an orbital decay timescale $\frac{P}{\dot{P}} = 3.25$\,Myr and finally to a modified stellar quality factor of $Q'_* = 1.8\times 10^5$, as found by \citet{2020ApJ...888L...5Y}. $Q'_*$ describes the efficiency with which orbital kinetic energy is dissipated within the star due to friction. The smaller its value, the stronger is the dissipation \citep{1966Icar....5..375G}.
There may also be a dependence of $Q'_*$ on the tidal forcing period $P_\mathrm{tide}$, as suggested by the results of \citet{penev2018}.\\

The observation of hot Jupiters enables us to gain information on the planet-star interactions within the system. We are examining tidal interactions between hot Jupiters and their host stars by measuring tidal orbital decay and also tidal deformation in the Guaranteed Time Observing (GTO) programme of the CHaracterising ExOplanet Satellite (CHEOPS) mission, under the "Tidal decay (ID 0012)" programme \citep{2021ExA....51..109B, Barros2022}. This paper deals with the tidal orbital decay candidates KELT-9\,b, discovered by \citet{2017Natur.546..514G}, KELT-16\,b, discovered by \citet{2017AJ....153...97O}, WASP-4\,b, discovered by \citet{2008ApJ...675L.113W}, and also the potential transit timing variation (TTV) system HD\,97658, discovered by \citet{2011ApJ...730...10H}. 
Most of the short period targets with the highest expected tidal decay rates were discovered by ground-based surveys like the one of the Kilodegree Extremely Little Telescope (KELT) and WASP.
The KELT survey \citep{2007PASP..119..923P} was aimed at finding close-in orbiting giant planets and concluded in 2020 after 14 years of observations and more than 20 discovered exoplanets\footnote{\label{fn:planet_counts}According to \href{http://exoplanet.eu}{exoplanet.eu} as of February 28, 2022.}. The WASP survey \citep{2006PASP..118.1407P} started in 2004 with SuperWASP-North, and WASP South joining in 2006. This survey discovered more than 180 exoplanets\footnote{See footnote \ref{fn:planet_counts}.}.
Our targets WASP-4\,b, KELT-9\,b, and KELT-16\,b are hot Jupiters with orbital periods of less than 1.5\,d and masses greater than 1.2 times that of Jupiter, making them fall into the category of potential orbital decay candidates. The suspected TTV candidate HD\,97658\,b is a super-Earth, which orbits its host star every 9.5\,d.\\

Measuring the orbital decay rate of our candidates allows us to constrain their $Q_*'$ values and to compare them against values from theory in the literature. Values given in the literature for the value of the modified stellar quality factor cover a broad range from $10^5$ up to $10^{8.5}$ (e.g. \citealt{Meibom2005,Jackson2008,penev2012}).\\

Amongst the space telescopes capable of observing exoplanet transits and occultations photometrically, especially two of them, CHEOPS and the Transiting Exoplanet Survey Satellite (TESS), are suited for our needs, while also offering the high precision necessary for measuring tidal decay.
The CHEOPS space mission \citep{2021ExA....51..109B} is an S-class mission in the science programme of the European Space Agency (ESA) and was launched on the 18th of December 2019. This space telescope is aimed at follow-up observations of planets transiting bright stars, delivering high precision photometry to improve system parameters where applicable. It consists of a single defocused 32\,cm telescope, which is used to observe a single target at a time. We use CHEOPS to acquire new photometric data for our targets, since it can provide precise timings from its transit and secondary eclipse observations \citep{Lendl2020, Borsato2021, Hooton2022, Deline2022, Brandeker2022, Barros2022}.\\
Besides CHEOPS, also TESS observations are used to get precise timings for our analysis. This satellite was launched on the 18th of April 2018. Unlike CHEOPS, TESS is a survey satellite, observing many targets at once during one of its observation sectors. It consists of four cameras, each with a 10.5\,cm entrance pupil diameter lens assembly in front of four CCD detectors \cite{2014SPIE.9143E..20R}, also making use of the transit method. There are more than 220 confirmed exoplanets discovered with TESS, with almost 5,800 project candidates to date\footnote{According to the \href{https://exoplanetarchive.ipac.caltech.edu/}{NASA Exoplanet Archive}, visited on June 30, 2022.} \citep{2013PASP..125..989A, 2021ApJS..254...39G}.\\

This paper presents observations to constrain the tidal decay of WASP-4\,b, KELT-9\,b, and KELT-16\,b and to investigate the apparent transit timing variations in the HD\,97658\,b system presented in \citet{2021MNRAS.tmp.3057M}. In Section~\ref{sec:obs}, we describe the observations from the literature and the new observations used for the analysis part of this work. The methods used in this paper are described in more detail in Section~\ref{sec:methods}, the results from our analysis are shown and discussed in Section~\ref{sec:results}, and the conclusions can be found in Section~\ref{sec:conclusion}.


\section{Observations\label{sec:obs}}

\subsection{Targets}
KELT-9 is a bright, fast-rotating star at the border of the B/A-types with an effective temperature of around 10,000\,K, and a mass of 1.98$\pm0.02$\,$M_\odot$, hosting an inflated hot Jupiter on a 1.48\,d polar orbit with a mass of $2.88\pm0.84$M$_\mathrm{J}$ and a radius of 1.89$^{+0.06}_{-0.05}$R$_\mathrm{J}$ \citep{2017Natur.546..514G, 2019A&A...627A.165H}. KELT-9\,b is highly-irradiated with a dayside temperature of more than 4000\,K. An important feature to be considered in this system is gravity darkening caused by the fast rotating nature of the star. The fast stellar rotation leads to the equatorial radius of the star being larger as compared to the polar radius due to centrifugal forces in the star. This causes an oblate shape, with higher surface temperatures at the poles and lower temperatures at the equator, modifying the shape of transits in light curves \citep{2009ApJ...705..683B, 2020AJ....160....4A, 2022AJ....163..122C}. KELT-9 was amongst the targets of a recent study by \citet{ivshina2022}, who found no evidence for orbital decay in this system.\\
KELT-16 is an F7V star with a brightness of V = 11.7 and an effective temperature of around 6250\,K. It has a mass of $1.21^{+0.04}_{-0.05}$M$_\odot$, a radius of $1.36^{+0.06}_{-0.05}$R$_\odot$ and hosts a hot Jupiter in a 0.97\,d orbit. The planet, KELT-16\,b, has a mass of $2.75^{+0.16}_{-0.15}$M$_\mathrm{J}$ and is inflated with a radius of $1.415^{+0.084}_{-0.067}$R$_\mathrm{J}$ \citep{2017AJ....153...97O}. Its equilibrium temperature is around 2450\,K. There have been several studies examining potential orbital decay including those of \citet{2018AcA....68..371M, 2020AJ....159..150P, 2021AJ....162..127W, 2022MNRAS.509.1447M} and \citet{ivshina2022}, with the conclusion of no significant period changes and hence no significant signs of orbital decay. 
A recent analysis by \citet{2022MNRAS.509.1447M} found a period derivative of $\dot{P} = -10.6\pm 13.1$\,ms\,yr$^{-1}$, consistent with a constant period, and hence no orbital decay.
From this, they were able to deduce a lower limit on the modified stellar tidal quality factor $Q'_* > (1.9 \pm 0.8) \times 10^5.$\\
WASP-4\,b is a 1.34\,d orbital period hot Jupiter with a mass of $1.186^{+0.090}_{-0.098}$M$_\mathrm{J}$, a radius of $1.321\pm0.039$R$_\mathrm{J}$, and an equilibrium temperature of around 1700\,K. This planet orbits a 5400\,K G7V star of V-magnitude 12.5 with a stellar mass of $0.97^{0.15}_{-0.09}$M$_\odot$ and a radius of $0.89^{+0.05}_{-0.06}$R$_\odot$, and was actually the first exoplanet discovered by the WASP South observatory \citep{2008ApJ...675L.113W, 2019AJ....157..217B}. For the WASP-4 system, there are several studies examining potential tidal orbital decay in this system, like those of \citet{2019AJ....157..217B, 2019MNRAS.490.1294B, 2019MNRAS.490.4230S, 2020ApJ...893L..29B, 2020MNRAS.496L..11B, 2021arXiv211209621T, 2022arXiv220211990M} and \citet{ivshina2022}. Of these, \citet{2020ApJ...893L..29B} find the orbital period of WASP-4\,b to be decreasing at a rate of $-8.64 \pm 1.26$\,ms\,yr$^{-1}$, but about $-6$\,ms\,yr$^{-1}$ of this being caused by a Doppler effect due to the system moving towards the Earth, as derived from radial velocity measurements. Later that year, \citet{2020MNRAS.496L..11B} examine this alleged radial velocity trend, but cannot recreate it using newer data. However, they find a decreasing period at $\dot{P} = -5.4\pm1.5$\,ms\,yr$^{-1}$ using previously unpublished data from \citet{2013MNRAS.434...46H} and \citet{2017AJ....154...95H}. \citet{2021arXiv211209621T} find a decay rate of $-7.33\pm0.71$\,ms\,yr$^{-1}$ using data from the latest TESS sectors for this target, and suspect an additional planet candidate with an orbital period of around $7000\,$d and a mass of approximately 5.5\,M$_\mathrm{J}$. The newest papers regarding this topic, from \citet{2022arXiv220211990M} and \citet{ivshina2022}, find a lower period decay rate of $\dot{P} = -4.8\pm 1.4$\,ms\,yr$^{-1}$ and $\dot{P} = -5.81\pm1.58$\,ms\,yr$^{-1}$ using data covering the same time as \citet{2021arXiv211209621T}, but excluding the \citet{2013MNRAS.434...46H} and \citet{2017AJ....154...95H} light curves for the former, and all available light curves including an additional Sector of TESS data for the latter.\\
The HD\,97658 system contains a K1V dwarf star with an effective temperature of around 5200\,K, stellar mass of $0.85\pm0.08\,$M$_\odot$ and a stellar radius of $0.728\pm0.008\,$R$_\odot$. It also contains a dense $7.62\pm0.42\,$M$_\oplus$, $2.293\pm0.07\,$R$_\oplus$ super-Earth, transiting every 9.49\,d \citep{2011ApJ...730...10H, 2021AJ....162..118E, 2021MNRAS.tmp.3057M}. At first, this planet was only detected using radial velocity data from \citet{2011ApJ...730...10H}, with \citet{Henry2011} claiming the observation of a transit, which turned out to not be a transit of the planet, since it could not be re-observed by \citet{2012ApJ...759L..41D} using the previously derived parameters a year later. However, using the MOST space telescope \citep{2003PASP..115.1023W} to observe HD\,97658 in 2012 and 2013, \citet{2013ApJ...772L...2D} were able to observe five transits of HD\,97658\,b at different times as those claimed by \citet{Henry2011} and confirm the transiting nature of this planet. The newest study including this system was carried out by \citet{2021MNRAS.tmp.3057M} using archival literature data and also a newly acquired CHEOPS transit observation. Using this data, they find unexplained apparent TTVs, resulting in a quadratic ephemeris giving a lower BIC value of 37.9 in comparison to 55.5 for the linear model. We examine this system with three additional CHEOPS visits and new TESS Sector 49 data to help us to differentiate between the models.\\

\subsection{New CHEOPS observations}\label{sec:obs_cheops}
For all of our targets, we provide at least one new observation made with the CHEOPS space telescope. In total, we obtained 19 new transit light curves with CHEOPS. 
Details regarding these can be found in the observation log in Table~\ref{tab:obs_log}. The ``OPTIMAL'' aperture was used for every light curve. The exposure times are 36.7\,s for KELT-9\,b, 60\,s for KELT-16\,b and WASP-4\,b, and 33\,s for HD\,97658\,b. The efficiency describes the percentage of time on the target spent collecting data. With CHEOPS, this is usually less than 100\,\% because its orbital configuration leads to the Earth blocking the line-of-sight to the target of the observation during parts of the satellite's orbit.
All of these light curves can be accessed via the DACE (Data \& Analysis Center for Exoplanets) website\footnote{\href{https://dace.unige.ch/cheopsDatabase/?}{https://dace.unige.ch/cheopsDatabase/?}} of the University of Geneva. Moreover, the light curves can also be downloaded and processed using \textsc{pycheops} \citep{2021MNRAS.tmp.3057M} and the file keys given in Table~\ref{tab:obs_log}.
All of the CHEOPS data was reduced with version 13.1.0 of the data reduction pipeline (DRP) \citep{2020A&A...635A..24H}.

\begin{table*}[h]
    \centering
    \renewcommand{\arraystretch}{1.1} 
    \caption{Observation log of new CHEOPS transit observations for all targets. }
    \begin{tabular}{c c c c c c c }
    \hline
        Target & Visit & Start date & Duration & No. of & Eff. & File key \\
                & no. & (UTC) & [h] & data points & [\%] & \\\hline
        KELT-9\,b & 1 & 2020-08-24 14:10:59 & 12.11 & 728 & 61.1 & CH\_PR100013\_TG001001\_V0200\\[4pt]
        KELT-16\,b & 1 & 2020-08-18 22:23:41 & 7.95 & 325 & 68.0 & CH\_PR100012\_TG001401\_V0200\\
            & 2 & 2020-09-06 07:44:45 & 8.75 & 364 & 69.2 & CH\_PR100012\_TG001402\_V0200\\
            & 3 & 2022-07-15 16:23:21 & 10.99 & 469 & 71.0 &  CH\_PR100012\_TG000501\_V0200\\
            & 4 & 2022-08-25 08:26:22 & 8.59 & 352 & 69.0 &  CH\_PR100012\_TG000502\_V0200\\
            & 5 & 2022-09-13 17:32:22 & 8.14 & 262 & 53.6 &  CH\_PR100012\_TG000503\_V0200\\
            & 6 & 2022-09-25 09:27:12 & 7.50 & 284 & 62.9 & CH\_PR100012\_TG000504\_V0200\\
            & 7 & 2022-10-01 03:59:13 & 8.14 & 282 & 57.6 & CH\_PR100012\_TG000505\_V0200\\[4pt]
        WASP-4\,b & 1 & 2021-07-15 10:19:20 & 5.80 & 203 & 58.1 & CH\_PR100012\_TG001101\_V0200\\
            & 2 & 2021-08-09 20:22:19 & 5.74 & 221 & 64.0 & CH\_PR100012\_TG001102\_V0200\\
            & 3 & 2021-08-17 20:21:21 & 6.35 & 235 & 61.5 & CH\_PR100012\_TG001103\_V0200\\
            & 4 & 2021-08-31 05:32:21 & 6.35 & 241 & 63.1 & CH\_PR100012\_TG001104\_V0200\\
            & 5 & 2021-09-08 06:20:21 & 6.25 & 234 & 62.2 & CH\_PR100012\_TG001105\_V0200\\
            & 6 & 2021-09-22 23:48:21 & 6.07 & 278 & 76.1 & CH\_PR100012\_TG001106\_V0200\\
            & 7 & 2021-10-14 09:58:20 & 6.05 & 227 & 62.3 & CH\_PR100012\_TG001107\_V0200\\
            & 8 & 2022-08.08 20:15:20 & 6.32 & 220 & 57.9 & CH\_PR100012\_TG001001\_V0200\\[4pt]
        HD\,97658\,b & 1 & 2021-03-20 06:54:54 & 10.12 & 732 & 66.2 & CH\_PR100012\_TG001901\_V0200\\
            & 2 & 2021-03-29 18:28:55 & 10.31 & 754 & 67.0 & CH\_PR100012\_TG001902\_V0200\\
            & 3 & 2022-01-27 10:16:56 & 10.05 & 625 & 56.9 & CH\_PR100012\_TG002201\_V0200\\
        \hline
    \end{tabular}
    \tablefoot{``Eff.'' stands for efficiency. The number of data points and the efficiency do not take data points flagged by the DRP into account.}
    \label{tab:obs_log}
\end{table*}

\subsection{New TESS observations}
Besides the new CHEOPS observations, we also made use of previously unpublished TESS data for
HD\,97658\,b from Sector 49 (20\,s cadence). 
In all cases in which we made use of data produced by the TESS Science Processing Operations Center (SPOC) at NASA Ames Research Center \citep{jenkinsSPOC2016}, except for HD97658b where we use the Simple Aperture Photometry (SAP) flux \citep{twicken:PA2010SPIE, morris:PA2020KDPH}, we made use of the Presearch Data Conditioning Simple Aperture Photometry (PDCSAP) flux for the light curve analysis \citep{Stumpe2012, Stumpe2014, 2012PASP..124.1000S}.
We chose to use the SAP flux for HD\,97658\,b, since one more transit is contained in this data set, as compared to the PDCSAP flux, where the last transit is also on a slope.
All light curves are available at the MAST\footnote{\href{https://mast.stsci.edu/portal/Mashup/Clients/Mast/Portal.html}{https://mast.stsci.edu/}} portal.

\subsection{Previously published observations}
\subsubsection{KELT-9\,b}
For the analysis of KELT-9\,b, we used the light curves for this target from \citet{2017Natur.546..514G}, who supplied them to us in private communication\footnote{These light curves are now also available at the \href{https://exofop.ipac.caltech.edu/tess/target.php?id=16740101}{TESS ExoFOP website}.}. In total, that are 23 primary transits and 7 secondary eclipses, observed from 2014 to 2015 using different ground-based telescopes and filters, details to be found in the filenames of the light curves at the ExoFOP website.
There are TESS data from Sectors 14, 15 and 41 available for this target at the MAST portal with a cadence of 120\,s, which have also been used in the study of \citet{ivshina2022}.
In addition to the transit described in Section~\ref{sec:obs_cheops}, several observations of this target were taken with the CHEOPS space telescope, with a total of nine observed occultations and four phase curves. These were observed from July to September 2020 and July to August 2021. In more detail, transits were observed on 2020 September 01, September 2, September 11, 2021 July 31, August 01, August 22, and August 24 in the phase curves with CHEOPS \citep{jones2022}.
Furthermore, there is one phase curve including one transit and two secondary eclipses available from Spitzer's Infrared Array Camera (IRAC) at 4.5\,$\mu$m, which was originally published in \citep{2020ApJ...888L..15M} and subsequently re-reduced by \citet{jones2022}. We use this latter version of the light curve with no further detrending. We homogeneously re-fitted every single one of these light curves.

\subsubsection{KELT-16\,b}\label{sec:obs_KELT-16}
For KELT-16\,b, we used 19 transit light curves from \citet{2017AJ....153...97O} which were obtained using the KELT-North Follow-up Network (KELT-FUN) between May and December 2015. These light curves were provided to us by the authors in private communication. The observations were made from ten different member observatories of KELT-FUN using different filter sets, details to be found in Table~2 of \citet{2017AJ....153...97O}. Some of those transits were observed simultaneously using different telescopes or different filters. There are two transits that were observed with two telescopes or filters each, two that were observed with four telescopes or filters each, and seven that were observed a single time.\\
We also used eleven transit light curves from \citet{2018AcA....68..371M} that are publicly available. These were acquired between November 2016 and October 2018 using the telescopes and filters stated in Table~1 of their publication.\\
Two more light curves were supplied to us by \citet{2020AJ....159..150P} in private communication. These were observed on 2017 June 10 and 11 with the 1.2\,m telescope at the Fred Lawrence Whipple Observatory located in Arizona using images from the KeplerCam detector and a Sloan $r'$-band filter.\\
Besides these, we also analysed the public\footnote{The ExoClock data can be accessed via the project \href{https://www.exoclock.space/}{homepage} or via at \href{https://osf.io/wna5e/}{osf.io}.} light curves from the ExoClock project's second data release \citep{2022ApJS..258...40K}. This data release contains 32 KELT-16\,b transit light curves observed with different telescopes and filters, acquired between July 2018 and November 2020. Details regarding the observation setups can be found at the project homepage under ``ExoClock Observations''.\\
Additionally, we used the 36 public transit light curves from \citet{2022MNRAS.509.1447M}. These light curves were obtained between June 2016 and June 2021, using the telescopes and filters stated in Table~1 of their publication. In total, 30 planetary transits were observed, with three of them having been observed with two telescopes, and one with three telescopes. TESS data for this target were also obtained from the MAST, including data from Sector 15 and 41, both of which were obtained with a cadence of 120\,s and already used in earlier publications.
For this target we also re-fitted all light curves homogeneously.\\

\subsubsection{WASP-4\,b}\label{sec:obs_WASP-4}
For WASP-4\,b, we made use of the mid-transit times from the homogeneously re-analysed light curves from \citet{2020MNRAS.496L..11B}. In their re-analysis, they process a total of 124 light curves, including those from an earlier paper \citep{2019MNRAS.490.1294B}, where they used data from \citet{2008ApJ...675L.113W}, \citet{2009AA...496..259G}, \citet{2011ApJ...733..127S}, \citet{2012A&A...539A.159N}, \citet{2013ApJ...779L..23P}, and amateur observations. Additionally, they used transit light curves from \citet{2009MNRAS.399..287S} which were re-assessed in \citet{2019MNRAS.490.4230S} because of potential clock errors together with new observations. Moreover, they also obtained previously non-public light curves from \citet{2013MNRAS.434...46H} and \citet{2017AJ....154...95H}, with the latter from transmission spectroscopy with GEMINI, offering high precision. Details about the observations can be found in Section~2 of each publication. The then available TESS Sector 2 light curve was used by them as well, and also six new amateur observations.\\
In our analysis, we used the available TESS light curves for this target from Sectors 2 (120\,s cadence), 28 and 29 (both 20\,s cadence), obtained from the MAST portal, and re-analysed them ourselves. Also, we made use of the publicly available light curves from the ExoClock project again, and re-fit these as well. Details about the observing setups of these can be found on the project homepage. In addition, we also obtained the original WASP light curves used in \citet{2008ApJ...675L.113W} from 2006, and also those recorded in the years 2007, 2010, and 2011, and re-fitted them.
Besides the transits, we used 4 secondary eclipse timings from the literature \citep{2011A&A...530A...5C, 2011ApJ...727...23B, 2015MNRAS.454.3002Z}, which were observed with ESO's VLT with the Infrared Spectrometer And Array Camera in the $K_S$-band, two with the Spitzer Space Telescope's IRAC camera (warm Spitzer) at 3.6\,$\mu$m and 4.5\,$\mu$m, and one with the Anglo-Australian telescope using the IRIS2 instrument in the $K_S$-band.

\subsubsection{HD\,97658\,b}
For the analysis of the HD\,97658 system, as \citet{2021MNRAS.tmp.3057M}, we also used the published mid-transit time of one Spitzer light curve at 4.5\,$\mu$m from \citet{2014ApJ...786....2V}, and 18 mid-transit times from \citet{2020AJ....159..239G} using HST/WFC3 spectroscopy, STIS on HST, the Spitzer Space Telescope at 3.6\,$\mu$m and 4.5\,$\mu$m, and the MOST Space Telescope in its 0.5\,$\mu$m bandpass. More details can be found in Section~2 and 3 in \citet{2020AJ....159..239G}.
Additionally, we also used the single mid-transit time from TESS given in \citet{2021MNRAS.tmp.3057M}, but not the time of mid-transit for the then only available CHEOPS light curve from April 2020. This was done to homogeneously re-analyse it together with three further light curves recorded by CHEOPS, two of which were observed in March 2021, with the remaining one having been observed in January 2022. 


\section{Transit and Secondary Eclipse Fitting\label{sec:methods}}

For the analysis of photometric CHEOPS data, we first use the \textsc{python} package \textsc{pycheops} \citep{2021MNRAS.tmp.3057M}. \textsc{pycheops} is a publicly available\footnote{Available at \href{https://github.com/pmaxted/pycheops}{GitHub}.} \textsc{python} module for the analysis of data from the ESA CHEOPS mission. This package can help to deal with the systematic effects present for the photometric data of this space telescope, induced by its nadir-locked orbit \citep{2021MNRAS.tmp.3057M}. Moreover, \textsc{pycheops} offers the ability to detrend the data of other effects, such as correlated noise using Gaussian process regression, fit transit and eclipse models to the light curves to retrieve certain parameters, and more. Fitting can be done using least-squares minimization and also Markov chain Monte Carlo algorithms. More in-depth information can be found in \citet{2021MNRAS.tmp.3057M}.
Besides the above and data visualization, this module also offers a built-in client for data handling, access and retrieval from the DACE website\footnote{\href{https://dace.unige.ch/}{https://dace.unige.ch/}} (Data \& Analysis Center for Exoplanets) hosted by the University of Geneva. General users can access the public data from CHEOPS, with CHEOPS science team members also having the possibility to access proprietary data.\\
We use this to download the data using the respective file keys. In all cases, we use the ``OPTIMAL'' aperture and also use the ``decontaminate'' option, which performs a subtraction of the contamination from nearby sources. We plot the light curve using \textsc{matplotlib} \citep{Hunter:2007} and perform a $2\sigma$ outlier clipping using the ``clip\_outliers'' function from \textsc{pycheops}. In the case of phase curves, we cut out the individual transits or occultations and separate them. Afterwards, we make use of \textsc{pycheops}' ``flatten'' function with the center value of the mask being the center of the transit or occultation from the plot. The mask width value is chosen to fit the transit or occultation width. 
Next, we save the individual transit or occultation light curves, consisting of time, flux, flux errors, and roll angle values into two separate files each, one including the roll angle information and the other one not. The data handling is being done using \textsc{numpy} \citep{2020Natur.585..357H}.
These two data files will be read in from the Transit and Light Curve Modeller (\textsc{tlcm}, version 97) \citep{2020MNRAS.496.4442C}. \textsc{tlcm} is a free software tool\footnote{It is freely available at \href{http://transits.hu}{transits.hu} together with a user manual.} to analyse, fit and simulate light curves and radial velocity curves of transiting exoplanets and detached eclipsing binaries. The \citet{2002ApJ...580L.171M} model is used to for the description of transits and occultations, and it is enhanced by also accounting for beaming, gravity darkening (see Section~\ref{sec:K9_transit_fitting}), reflection and ellipsoidal effects of both the star and the planet. It also features a wavelet model for red noise, based on the model of \citet{2009ApJ...704...51C}. \textsc{tlcm} has the ability to fit photometric, as well as radial velocity data simultaneously, including a simplified Rossiter-McLaughlin effect. A genetic algorithm is used to find the global minimum of the $\chi^2$ or $\log L$ values. Afterwards, a simulated annealing algorithm is used to refine the fit, with a Markov chain Monte Carlo algorithm being used for error estimation.\\
The light curve files are read in by \textsc{tlcm} by entering their names into the configuration file and into a script handling the roll angle decorrelation using six parameters to fit the roll angle effect to the data set using the following formula:
\begin{align*}
    f_\mathrm{RA} = \,&p_1 \sin(1\cdot RA) + p_2\sin(2\cdot RA) + p_3\sin(3\cdot RA)\,+\\ &p_4\cos(1\cdot RA) + p_5\cos(2\cdot RA) + p_6\cos(3\cdot RA),
\end{align*}
with $f_\mathrm{RA}$ being the value needed to add to the corresponding flux value to obtain the decorrelated flux, $p_1$ to $p_6$ being the six fit parameters, and $RA$ being the respective roll angle value in radians. The roll angle decorrelation could, however, also be done in \textsc{pycheops}.\\
The necessary fit parameters are the semi-major axis of the planetary orbit, planet-to-star radius ratio, impact parameter, limb darkening parameters, epoch of the transit, and the orbital period if multiple transits are fitted at the same time. These values, except for the limb darkening parameters, were taken from TEPCAT \citep{2011MNRAS.417.2166S.TEPCAT} as starting points for each fit, except for the transit epoch, which has been read off of the transit plots for each transit. 
After running \textsc{TLCM}, we extracted the transit epochs, as well as their error bars from the results files, using the median solution values.
In the cases where we created combined models using several transits e.g. from TESS, we did a \textsc{TLCM} run while fitting for the parameters as stated above, using the epoch of one of the transits near the center of the data set. After completion of the run, we use the resulting parameters values for the individual transits and fix them, except for the epoch, which is determined visually for each transit as before. As in the case of individual transit fits without a combined model, the resulting epochs are then used in the TTV fits of our three models.\\

For the occultation fitting, we use the same procedure as before to cut out the occultations from the light curves. 
Afterwards, we phase-fold the eclipses from one observing season, except for Spitzer data, using the \textsc{PyAstronomy} package \citep{pya} and fit them using a \textsc{batman} \citep{2015PASP..127.1161K} box model together with \textsc{lmfit} \citep{newville_matthew_2014_11813} and \textsc{emcee} \citep{2013PASP..125..306F} for the Markov chain Monte Carlo (MCMC) fitting. Our box model uses an epoch near the center of the data set, together with the period, planet radius, semi-major axis, inclination, eccentricity, and uniform limb darkening. The values for these parameters, as before, were taken from TEPCAT. First, we do a preliminary fit to get the combined shape of all occultations in the data set, by leaving all parameters free with reasonable error bars and the best-fit period from the linear model of the TTV fit for each target. After this, we fix the eclipse shape to the resulting values of the fit and leave only the epoch as a free parameter. Using 5000 steps, 500 steps of burn-in and 200 walkers for the MCMC algorithm of \textsc{emcee}, we fit for the epoch and its error. The resulting value is then used in the corresponding TTV fit.

\section{Timing Analysis and Results}\label{sec:results}
\subsection{Timing analysis models}\label{sec:models}
\subsubsection{Keplerian orbit model}
To analyse the timing data, we used three models, following \citet{2017AJ....154....4P}. The first model assumes a Keplerian system with a circular orbit, a constant orbital period and hence a linear ephemeris:
\begin{align}
    t_\mathrm{tra}(N) &= t_0 + N \, P, \\
    t_\mathrm{occ}(N) &= t_0 + \frac{P}{2} + N \, P,
\end{align}
with $t_\mathrm{tra}$ and $t_\mathrm{occ}$ the calculated mid-transit and mid-occultation times, $t_0$ the reference mid-transit time, $N$ the number of orbits from the reference mid-transit time, and $P$ the orbital period of the exoplanet. The parameters $t_0$ and $P$ are fitted in this model.\\

\subsubsection{Orbital decay model}
The second model is quadratic and assumes a circular orbit with a constant change in the orbital period due to angular momentum transfer from the planet to the star (see e.g. \citealt{1973ApJ...180..307C, 1996ApJ...470.1187R}):
\begin{align}
    t_\mathrm{tra}(N) &= t_0 + N \, P + \frac{1}{2}\,\frac{dP}{dN}\,N^2, \\
    t_\mathrm{occ}(N) &= t_0 + \frac{P}{2} + N \, P + \frac{1}{2}\,\frac{dP}{dN}\,N^2.
\end{align}
In this model, we fit for $t_0$, $P$, and the decay rate $\frac{dP}{dN}$ which is related to the period derivative $\dot{P}$ by $\dot{P} = \frac{dP}{dt} = \frac{1}{P}\,\frac{dP}{dN}$.\\
However, if the planet is on an inclined orbit around its host star, this can have an effect on the rate of the tidal decay.
According to Kaula's theory of tide \citep[see e.g.][Eq.~(138)]{Boue19}, the evolution of the semimajor axis of a circular orbit is, at the quadrupole order $l=2$,
\begin{equation}
\begin{aligned}
\frac{\mathrm{d}a}{\mathrm{d}t} = -& 2\,a\,n\,\frac{M_p}{M_*}\left(\frac{R_*}{a}\right)^5\sum_{m=0}^l \frac{(l-m)!}{(l+m)!} (2-\delta_{0m})\\ &\cdot \sum_{p=0}^l (l-2p) F_{lmp}^2(\psi) K_2(\omega_{lmp0})\,,
\label{eq.dadt}
\end{aligned}
\end{equation}
with $\omega_{lmpq} \approx (l-2p+q)n-m\,\dot\theta$ where $n$ is the orbital mean motion and $\dot\theta$ the angular speed of the star. In the semimajor axis variation rate (Eq.~\ref{eq.dadt}), $F_{lmp}(\psi)$ are the inclination functions defined in \citep{Kaula64} and $K_2$ is the quality function of the star. Under the constant phase lag model, the expression of $K_2$ in terms of the modified quality factor $Q'_*$ is
\begin{equation}
K_2(\omega) = \frac{3}{Q'_*}\mathrm{Sign}(\omega)\,.
\label{eq.K2}
\end{equation}
All the tidal frequencies $\omega_{lmpq}$ have a constant sign except $\omega_{2100} = 2n-\dot\theta$ and $\omega_{2200} = 2n-2\dot\theta$. Therefore, in the calculation of $\dot a$ (\ref{eq.dadt}), 5 cases have to be considered, namely, $\dot\theta < n$, $\dot\theta = n$, $n < \dot\theta < 2n$, $\dot\theta = 2n$, and $\dot\theta > 2n$. In terms of the periods $P = 2\pi/n$ and $P_* = 2\pi/\dot\theta$, the expression of the orbital period derivative $\dot P/P = \frac{3}{2}\dot a/a$ is
\begin{equation}
\dot P = f\frac{\pi}{Q'_*}\left(\frac{M_p}{M_*}\right)\left(\frac{R_*}{a}\right)^5\,,
\label{eq.Pdot}
\end{equation}
with $f$ a numerical factor given in Table~\ref{tab.tidalfactor}. In particular, for a short period planet $P < P_*$, we retrieve the formula of \citet{1966Icar....5..375G} i.e.
\begin{equation}\label{eq:qstar}
\left(\dot P\right)_{P<P_*} = -\frac{27\,\pi}{2\,Q'_*}\left(\frac{M_p}{M_*}\right)\left(\frac{R_*}{a}\right)^5\,.
\end{equation}

\begin{table*}[t]
\caption{\label{tab.tidalfactor}Tidal factor $f$ in the expression of the period derivative (\ref{eq.Pdot}) as a function of the orbital obliquity $\psi$, that is the angle between the orbital and stellar rotational plane, reckoned from the stellar equator, for an equatorial orbit EO ($\psi=0$), and for a polar orbit PO ($\psi=90^\circ$).}
\begin{center}
\renewcommand{\arraystretch}{1.4}
\begin{tabular}{llcc}
\hline
case & general expression & EO & PO \\ \hline
$P < P_*$ & $-\frac{27}{2}$ & $-\frac{27}{2}$ & $-\frac{27}{2}$\\
$P = P_*$ & $-\frac{27}{4}+\frac{27}{4}\cos(\psi)-\frac{27}{4}\sin^2(\psi)-\frac{27}{8}\cos(\psi)\sin^2(\psi)+\frac{27}{32}\sin^4(\psi)$ & $0$ & $-\frac{405}{32}$\\
$P_* < P < 2P_*$ & $\frac{27}{2}\cos(\psi)-\frac{27}{2}\sin^2(\psi)-\frac{27}{4}\cos(\psi)\sin^2(\psi)+\frac{27}{16}\sin^4(\psi)$ & $+\frac{27}{2}$ & $-\frac{189}{16}$\\
$P = 2P_*$ & $\frac{27}{2}\cos(\psi)-\frac{27}{4}\sin^2(\psi)-\frac{27}{16}\sin^4(\psi)$ & $+\frac{27}{2}$ & $-\frac{135}{16}$\\
$P > 2P_*$ & $\frac{27}{2}\cos(\psi)+\frac{27}{4}\cos(\psi)\sin^2(\psi) - \frac{81}{16}\sin^4(\psi)$ & $+\frac{27}{2}$ & $-\frac{81}{16}$\\ \hline
\end{tabular}
\end{center}
\end{table*}

This enables us to estimate the modified tidal quality factor $Q'_*$ from the measured rate of orbital decay. This rate can also provide us with a lower limit on $Q'_*$, which can be obtained by calculating $Q'_*$ using the 95\% confidence lower limit of $\dot{P}$, with the errors being derived from propagating the uncertainties in $M_p/M_*$ and $R_*/a$, as has been done in e.g. \citet{2020AJ....159..150P} and \citet{2022MNRAS.509.1447M}.

\subsubsection{Apsidal precession model}
The third model assumes a non-zero eccentricity $e$ and apsidal precession, following the formulations of \citet{1995Ap&SS.226...99G}:
\begin{align}
    t_\mathrm{tra}(N) &= t_0 + N \, P_s - \frac{e\,P_a}{\pi}\cos\omega(N), \\
    t_\mathrm{occ}(N) &= t_0 + \frac{P_a}{2} + N \, P_s + \frac{e\,P_a}{\pi}\cos\omega(N),
\end{align}
with $P_s$ the sidereal period, $P_a$ the anomalistic period, and $\omega$ the argument of pericenter. The dependency of $\omega$ on $N$ is as follows:
\begin{equation}
    \omega(N) = \omega_0 + \frac{d\omega}{dN}\,N,
\end{equation}
with $\omega_0$ the argument of pericenter at the reference time. The relation of the sidereal period and the anomalistic period can be described as:
\begin{equation}
    P_s = P_a \left( 1-\frac{1}{2\pi}\frac{d\omega}{dN} \right).
\end{equation}
In this model, we fit for the parameters $t_0$, $P_s$, $e$, $\omega_0$, and $\frac{d\omega}{dN}$.\\
Should there be an eccentricity in the system, we would expect it to be damped by the tidal forces. Both the tides raised on the star and on the synchronous planet tend to circularize the orbit. From Kaula's theory of tides \citep[e.g.][Eq.~154]{Boue19}, the rate of eccentricity damping, computed at first order in eccentricity, can be put in a form similar to that in \citet{Yoder81}, namely,
\begin{equation}
\frac{\mathrm{d}e}{\mathrm{d}t} = -\frac{1}{3}\, c\, (7 D + f_e)\,e\,,
\label{eq.edot}
\end{equation}
where the constants $c$ and $D$ are given by
\begin{equation}
c = \frac{27}{2Q'_\star} \frac{M_p}{M_\star} \left(\frac{R_\star}{a}\right)^5 n
\quad\mathrm{and}\quad
D = \frac{k_2}{3} \left(\frac{R_p}{R_\star}\right)^5 \left(\frac{M_\star}{M_p}\right)^2 \frac{Q'_\star}{Q_p}\,.
\end{equation}
In these formulae, $k_2$ is the planet's second Love number, $Q_p$ its tidal quality factor and $R_p$ its radius. The parameter $f_e$, provided in Table~\ref{tab.tidalfactore}, represents the contribution of the tides raised on the star to the circularization.

\begin{table*}
\caption{\label{tab.tidalfactore}Tidal factor $f_e$ in the expression of the eccentricity derivative (Eq.~\ref{eq.edot}) as a function of the orbital obliquity $\psi$, the angle between the orbital and stellar rotational plane, reckoned from the stellar equator, for an equatorial orbit EO ($\psi=0$), and for a polar orbit PO ($\psi=90^\circ$).}
\begin{center}
\renewcommand{\arraystretch}{1.4}
\begin{tabular}{l*{5}{r}cc}
\hline
case & \multicolumn{5}{l}{general expression} & EO & PO \\ \hline
$P < 1/2\,P_\star$ & $+\frac{25}{4}$ &&&&& $+\frac{25}{4}$ & $+\frac{25}{4}$\\
$P = 1/2\,P_\star$ & $
 + \frac{101}{16}                        $&$\!\!\!\!
 + \frac{1}{16}\cos(\psi)                        $&$\!\!\!\!
 - \frac{1}{16}\sin^2(\psi)                        $&$\!\!\!\!
 - \frac{1}{32}\sin^2(\psi)\cos(\psi)                 $&$\!\!\!\!
 - \frac{35}{128}\sin^4(\psi)$ & $+\frac{51}{8}$ & $+\frac{765}{128}$\\
$1/2\,P_\star < P < P_\star$ & $
 + \frac{51}{8}                        $&$\!\!\!\!
 + \frac{1}{8}\cos(\psi)                        $&$\!\!\!\!
 - \frac{1}{8}\sin^2(\psi)                        $&$\!\!\!\!
 - \frac{1}{16}\sin^2(\psi)\cos(\psi)                 $&$\!\!\!\!
 - \frac{35}{64}\sin^4(\psi)$ & $ +\frac{13}{2} $ & $+\frac{365}{64}$ \\
$P = P_\star$ & $
 + \frac{53}{8}                        $&$\!\!\!\!
 + \frac{3}{8}\cos(\psi)                        $&$\!\!\!\!
 - \frac{23}{16}\sin^2(\psi)                        $&$\!\!\!\!
 - \frac{1}{8}\sin^2(\psi)\cos(\psi)                 $&$\!\!\!\!
 + \frac{37}{64}\sin^4(\psi)$ & $+7$ & $+\frac{369}{64}$\\
$P_\star < P < 3/2\,P_\star$ & $
 + \frac{55}{8}                        $&$\!\!\!\!
 + \frac{5}{8}\cos(\psi)                        $&$\!\!\!\!
 - \frac{11}{4}\sin^2(\psi)                        $&$\!\!\!\!
 - \frac{3}{16}\sin^2(\psi)\cos(\psi)                 $&$\!\!\!\!
 + \frac{109}{64}\sin^4(\psi)$ & $+\frac{15}{2}$ & $+\frac{373}{64}$\\
$P = 3/2\,P_\star$ & $
 + \frac{61}{16}                        $&$\!\!\!\!
 - \frac{39}{16}\cos(\psi)                        $&$\!\!\!\!
 + \frac{5}{16}\sin^2(\psi)                        $&$\!\!\!\!
 + \frac{43}{32}\sin^2(\psi)\cos(\psi)                 $&$\!\!\!\!
 + \frac{169}{128}\sin^4(\psi)$ & $+\frac{11}{8}$ & $+\frac{697}{128}$\\
$3/2\,P_\star < P < 2P_\star$ & $
 + \frac{3}{4}                        $&$\!\!\!\!
 - \frac{11}{2}\cos(\psi)                        $&$\!\!\!\!
 + \frac{27}{8}\sin^2(\psi)                        $&$\!\!\!\!
 + \frac{23}{8}\sin^2(\psi)\cos(\psi)                 $&$\!\!\!\!
 + \frac{15}{16}\sin^4(\psi)$ & $-\frac{19}{4}$ & $+\frac{81}{16}$\\ 
$P = 2P_\star$ & $
 + \frac{3}{4}                        $&$\!\!\!\!
 - \frac{11}{2}\cos(\psi)                        $&$\!\!\!\!
 + \frac{29}{8}\sin^2(\psi)                        $&$\!\!\!\!
 + \frac{25}{8}\sin^2(\psi)\cos(\psi)                 $&$\!\!\!\!
 + \frac{13}{16}\sin^4(\psi)$ & $-\frac{19}{4}$ & $+\frac{83}{16}$\\
$2P_\star < P < 3 P_\star$ & $
 + \frac{3}{4}                        $&$\!\!\!\!
 - \frac{11}{2}\cos(\psi)                        $&$\!\!\!\!
 + \frac{31}{8}\sin^2(\psi)                        $&$\!\!\!\!
 + \frac{27}{8}\sin^2(\psi)\cos(\psi)                 $&$\!\!\!\!
 + \frac{11}{16}\sin^4(\psi)$ & $-\frac{19}{4}$ & $+\frac{85}{16}$\\
$P = 3P_\star$ & $
 + \frac{3}{4}                        $&$\!\!\!\!
 - \frac{11}{2}\cos(\psi)                        $&$\!\!\!\!
 + \frac{13}{16}\sin^2(\psi)                        $&$\!\!\!\!
 + \frac{5}{16}\sin^2(\psi)\cos(\psi)                 $&$\!\!\!\!
 + \frac{71}{32}\sin^4(\psi)$ & $-\frac{19}{4}$ & $+\frac{121}{32}$\\
$P > 3P_\star$ & $
 + \frac{3}{4}                        $&$\!\!\!\!
 - \frac{11}{2}\cos(\psi)                        $&$\!\!\!\!
 - \frac{9}{4}\sin^2(\psi)                        $&$\!\!\!\!
 - \frac{11}{4}\sin^2(\psi)\cos(\psi)                 $&$\!\!\!\!
 + \frac{15}{4}\sin^4(\psi)$ & $-\frac{19}{4}$ & $+\frac{9}{4}$\\
\hline
\end{tabular}
\end{center}
\end{table*}

\subsection{Results}
\subsubsection{KELT-9\,b}\label{sec:KELT-9_disc}

\subsubsection*{Transit fitting}\label{sec:K9_transit_fitting}
For the transit light curves publicly available from \citet{2017Natur.546..514G}, we fitted each transit individually with \textsc{tlcm}.
The reason for this is that the observations were made using different telescopes and filters, making it too computationally expensive to create a combined model from all light curves at once and achieve convergence, because of all of the extra parameters associated with the implementation of a different noise model per instrument and also different limb darkening parameters per filter. 
On the other hand, for the TESS and CHEOPS data, a combined model was created by jointly fitting both data sets with \textsc{tlcm} to improve the transit fits especially for the CHEOPS observations, since some of those lack parts of the ingress or egress, leading to uncertain fits. This is countered by fixing the transit shape from the combined model in the individual transit fits. We do not use priors for these, since our approach leads to the same mid-transit times as when using priors (within 1.5\,s), but gives less strict error bars by about 15\% to 20\%.
For this system, CHEOPS and TESS have relatively close theoretical limb darkening parameters with $a_\mathrm{TESS}=0.1712$, $b_\mathrm{TESS}=0.2399$, $a_\mathrm{CHEOPS}=0.2571$, and $b_\mathrm{CHEOPS}=0.3264$ \citep{2018A&A...618A..20C, 2021RNAAS...5...13C}. 
The validity of the combination of the two data sets to create a combined model is given by the comparison of the resulting transit times of the TESS and CHEOPS combined model against those of the combined TESS model from all TESS transit observations for this target, with a maximum difference in the mid-transit times of $3.6\,$s and most of them being smaller than $2\,$s.
An additional feature of the transit light curves for this target is gravity darkening, which can distort the transit shape and needs extra modelling to account for (see e.g. \citet{2020AJ....160....4A}). However, gravity darkening is implemented in \textsc{tlcm}, so we accounted for this in the creation of the combined model from CHEOPS and TESS and fixed it afterwards when fitting for the individual mid-transit times. 

\begin{figure}[!h]
    \centering
    \includegraphics[width=0.93\linewidth]{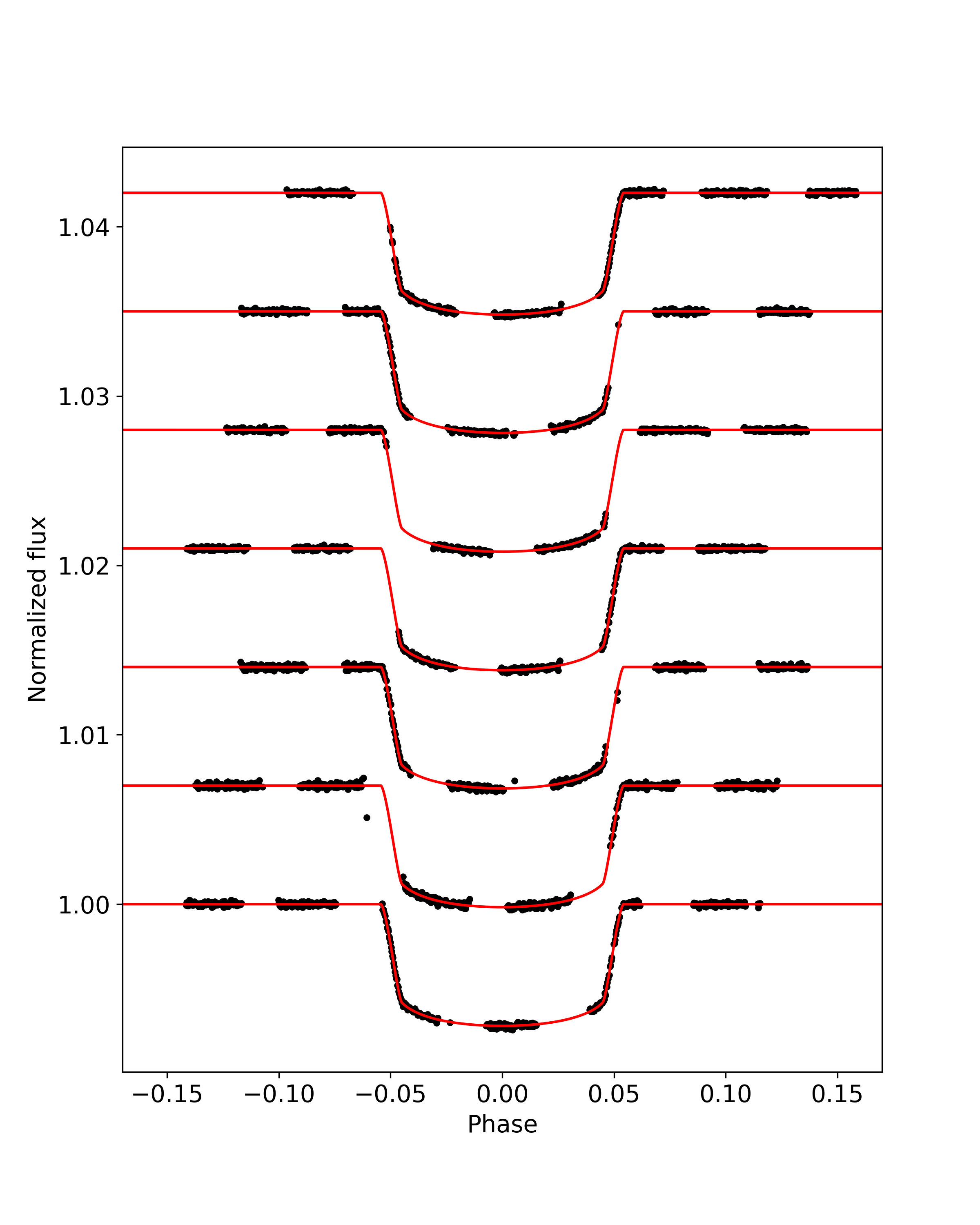}
    \caption{Transits of KELT-9\,b observed with CHEOPS. The x-axis shows the orbital phase, with the y-axis showing the normalized flux from the star. There is an offset of 0.007 in terms of normalized flux between the individual transits. The black dots represent the data, which were corrected for the roll angle of the satellite using \textsc{pycheops} and fitted with \textsc{tlcm}, leading to the transit models (red lines). The lowest transit resulted from the single transit observation listed in Table~\ref{tab:obs_log}, the rest were observed in phase curves \citep{jones2022}.}
    \label{fig:KELT-9_transits}
\end{figure}

TLCM parameterises the gravity-darkening by fitting for two angles, the inclination of the stellar rotation axis, and the angle between it and celestial north, $\Omega_*$ which is related to the sky-projected obliquity, $\lambda$ by $\Omega_* = 90^\circ - \lambda$ \citep{Lendl2020}. The gravity-darkening coefficient, $\beta$, is calculated from the relation between stellar oblateness and $\beta$ proposed by \citet{2011A&A...533A..43E}; for KELT-9, $\beta = 0.22$. The previously unpublished transit observed with CHEOPS is shown in Fig.~\ref{fig:KELT-9_transits} together with the already published transits, and the resulting mid-transit times are given in Table~\ref{tab:KELT-9_tts}.\\

\begin{table*}[h]
    \centering
    \setlength{\tabcolsep}{10pt} 
    \renewcommand{\arraystretch}{1.05} 
    \caption{KELT-9\,b mid-transit times and errors.}
    \begin{tabular}{c c c c}
        \hline
        Mid-transit time & Error & Epoch & Source\\
        $[\mathrm{BJD}_\mathrm{TDB}]$ & [d] & & \\\hline
        2456873.51672 &	0.00119 & -870 &	\citet{2017Natur.546..514G} \\
        2456876.47963 &	0.00117 & -868 &	\citet{2017Natur.546..514G} \\
        2456876.47981 &	0.00062 & -868 &	\citet{2017Natur.546..514G} \\
        2456895.73193 &	0.00134 & -855 & \citet{2017Natur.546..514G} \\
        2456895.73416 &	0.00151 & -855 & \citet{2017Natur.546..514G} \\
        \dots & \dots & \dots & \dots\\
        \hline
    \end{tabular}
    \tablefoot{Epoch given from the middle of the data set. The ``source'' column describes the source of the respective light curve. Only a portion of the data is shown here, the full table will be made available with the online version of the paper and at the CDS.}
    \label{tab:KELT-9_tts}
\end{table*}

\subsubsection*{Secondary eclipse fitting}
The occultation obervations for this target yield a total of eight occultations times. One from the ground-based observations of \citet{2017Natur.546..514G} with a relatively large error bar, two from Spitzer, three from TESS, and two from CHEOPS. The nature of the Spitzer data for this target allows the individual secondary eclipses of this target to be fitted with a similar precision to the transit fits, which is not the case for TESS and CHEOPS, here only the combination of multiple occultation observations allows a similar precision. Even though the precision of the CHEOPS secondary eclipse light curves would be good enough for individual occultation fits, the missing data points in between CHEOPS observations, when the target cannot be observed due to the Earth being in the line of sight of the target, make it hard to get precise timings, especially if parts of the ingress or egress are missing.

When mid-occultation times are derived from multiple occultation light curves, the reference time of conjunction is chosen to be as close to the middle of the data set as possible. This is done because the occultation observations are relatively evenly spaced in each data set.
The resulting mid-occultation times can be found in Table~\ref{tab:KELT-9_ots}.

\begin{table*}[h]
    \centering
    \setlength{\tabcolsep}{10pt} 
    \renewcommand{\arraystretch}{1.05} 
    \caption{KELT-9\,b mid-occultation times and errors.}
    \begin{tabular}{c c c c}
        \hline
        Mid-occultation time & Error & Epoch & Source\\
        $[\mathrm{BJD}_\mathrm{TDB}]$ & [d] & & \\\hline
        2457117.15824 &	0.00138 & -705.5 & \citet{2017Natur.546..514G}\\
        2458414.62225 &	0.00031 & 170.5 & \citet{2020ApJ...888L..15M}\\
        2458416.10255 &	0.00028 & 171.5 & \citet{2020ApJ...888L..15M}\\
        2458694.55498 &	0.00044 & 359.5 & TESS\\
        2458725.65775 &	0.00061 & 380.5 & TESS\\
        \dots & \dots & \dots & \dots\\
        \hline
    \end{tabular}
    \tablefoot{The epochs are given relative to the mid-time of the data set. The ``source'' column denotes the source of the respective light curves. Only a portion of the data is shown here, the full table will be made available with the online version of this paper and at the CDS.}
    \label{tab:KELT-9_ots}
\end{table*}

\subsubsection*{Timing analysis}\label{sect:fit_descr}
\begin{figure}[h]
    \centering
    \includegraphics[width=\linewidth]{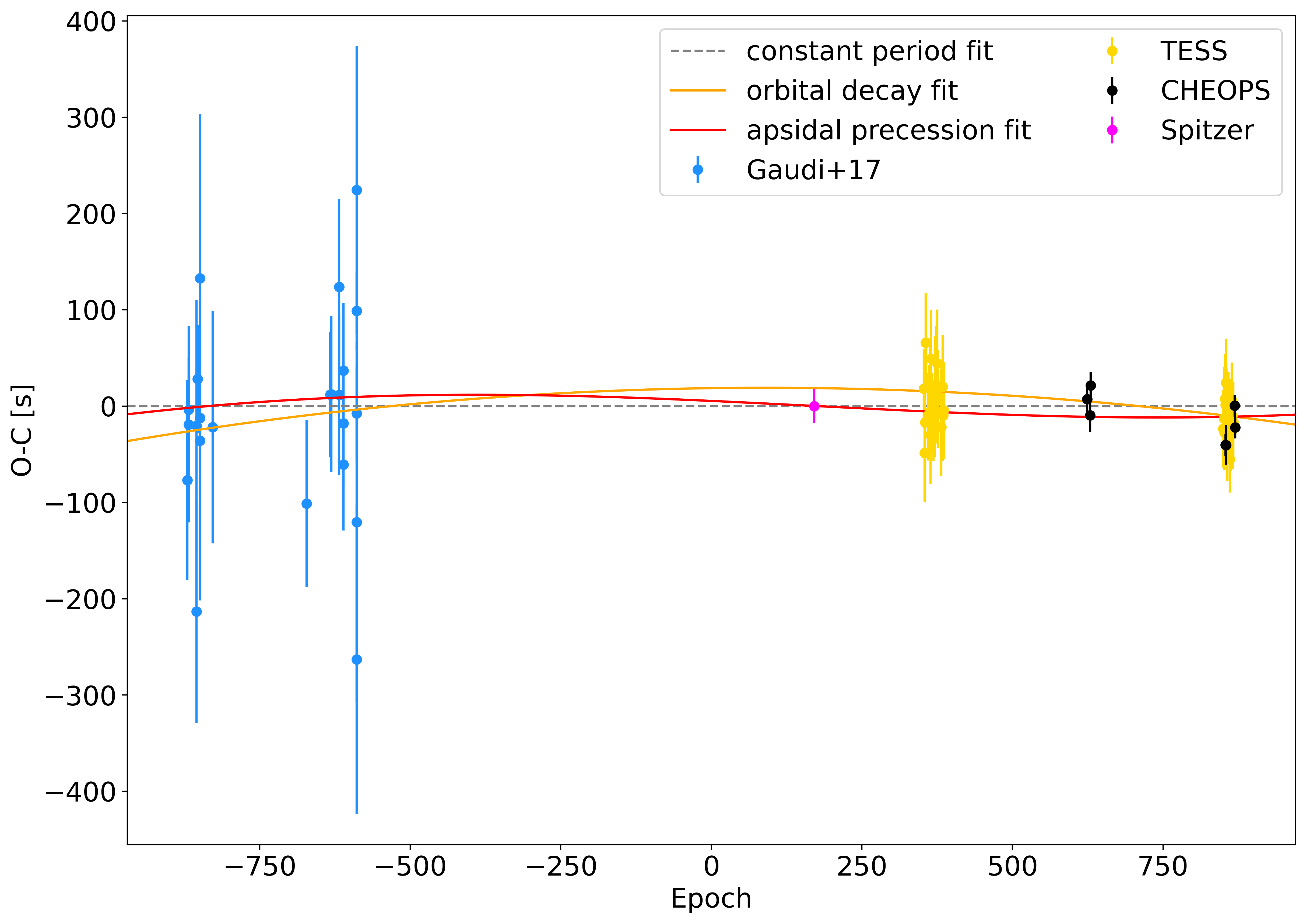}
    \caption{O-C plot showing the deviations in transit time from the best-fit linear ephemeris (gray dashed line) for KELT-9\,b. The transit number is shown on the x-axis. The y-axis shows the difference in observed and calculated mid-transit time. 
    The orange line shows the best orbital decay fit to the KELT-9\,b transit timing variation data, with the red line showing the best apsidal precession fit. CHEOPS data are highlighted in black.}
    \label{fig:Kelt-9_ttvs}
\end{figure}

\begin{figure}[h]
    \centering
    \includegraphics[width=\linewidth]{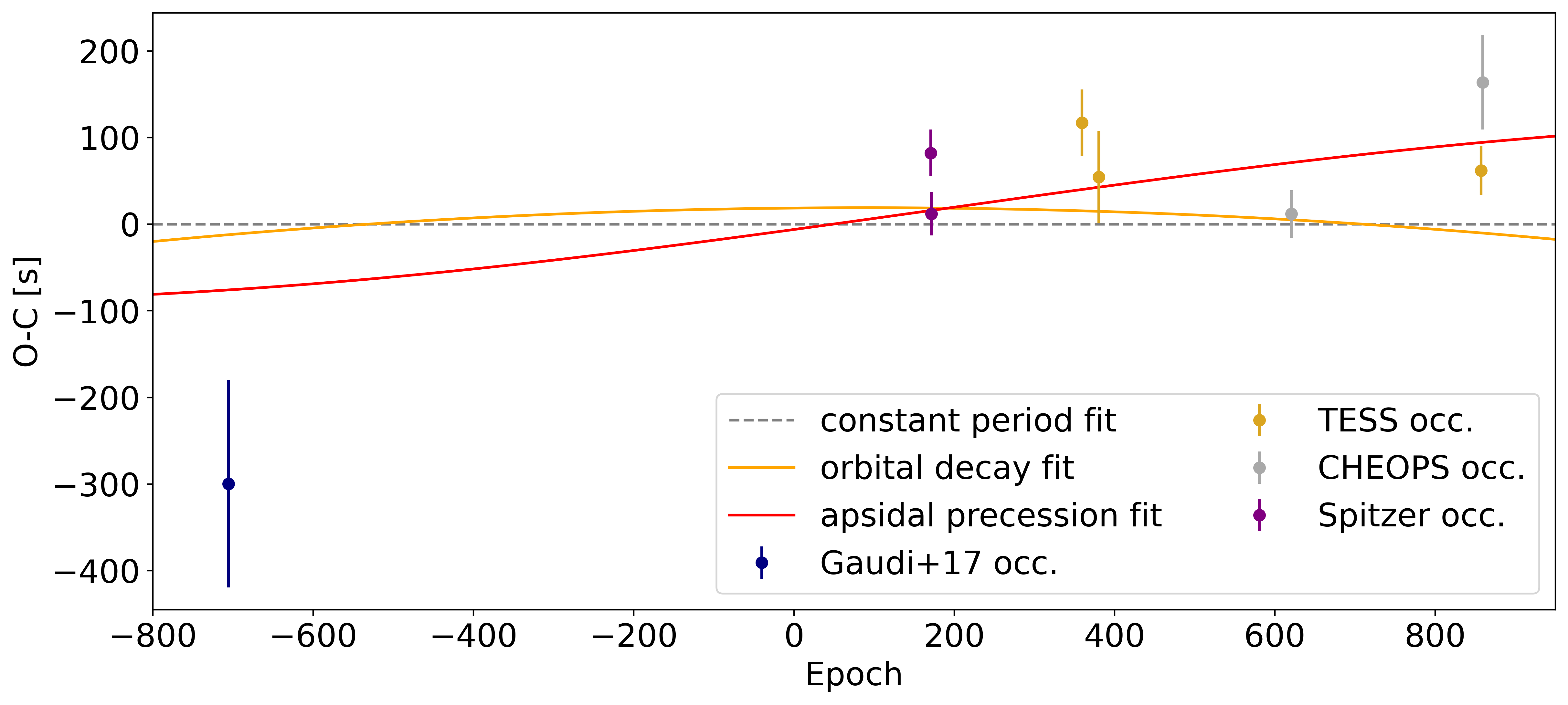}
    \caption{KELT-9\,b occultations from various sources, as stated in the legend. The x-axis shows the epoch of the occultation, and the y-axis shows the difference in observed and calculated mid-occultation time, assuming a linear ephemeris. The best fits to the constant period, orbital decay, and apsidal precession models are shown as the gray dashed, orange solid, and red solid lines.}
    \label{fig:KELT-9_occs}
\end{figure}

In combination, the transit and occultation data yield 76 times of mid-transit or mid-occultation, with the timings from the ground-based observations of \citet{2017Natur.546..514G} having relatively large error bars in comparison to the those of CHEOPS, Spitzer and TESS. 
These data points are fitted using the three models described in Section~\ref{sec:models} using \textsc{emcee}.
First of all, we gather the necessary planetary and stellar parameters from TEPCAT. Then, the three models are fitted to the transit and occultation timing data using an MCMC algorithm with 30\,000 steps, a burn-in period of 10\,000 steps, and 500 walkers. The data points are weighted according to their timing errors. 
The resulting fits are shown in Fig.~\ref{fig:Kelt-9_ttvs} and Fig.~\ref{fig:KELT-9_occs}, and the fit parameters for each model are available in Table~\ref{tab:fit_parameters}.\\

The BIC values indicate the preference of the apsidal precession model with $BIC_p = 90.5$ over the orbital decay ($BIC_d = 95.3$) and the Keplerian orbit model ($BIC_k = 103.7$), minding the number of free parameters of each fit. Furthermore, the $d\omega/dN$ value is in agreement with a constant period only at over $5\,\sigma$, and the resulting orbital period deviates from the other two at around $3\,\sigma$. The best-fit eccentricity value agrees with a circular orbit only at $4\,\sigma$. Furthermore, \citet{2022arXiv220302546S} recently found a nodal precession trend using radial velocity measurements, which could also have an effect on the transit timings, albeit generally a much smaller signal in comparison to apsidal precession for typical hot Jupiters. Transit duration variations are a better indicator for nodal precession \citep{miralda-escude2002, ragozzine2009, damiani2011}.\\
Moreover, the fits to the Spitzer data yield small error bars and are also in agreement with a constant period, which has an influence especially on the apsidal precession fit. The early data does not help in constraining the fit to either of the models since the scatter and also the error bars are large in comparison to the newer data sets. The occultation data from the various sources is also inconclusive in differentiating the three models because the scatter of the data points is too large (see Fig.~\ref{fig:KELT-9_occs}).\\
More high-precision observations of this target in the future are necessary to constrain the models further and help to differentiate between them. KELT-9 is observed in TESS Sector 55 in camera 3, from 2022 August 05 to September 01, which could enable better fits and might replace the apsidal precession model as the best model, since this is heavily influenced by the Spitzer transit data point.\\

Using the results from the orbital decay fit, we can calculate $\dot{P} = (-24.42\pm10.66)$ ms\,yr$^{-1}$, assuming the observed trend is true and only caused by this effect. Calculating the $95\%$ confidence lower limit on the orbital decay timescale $\tau = \frac{P}{|\dot{P}|}$ yields $\tau > (2.8\pm0.7)\,\mathrm{Myr.}$\\
Due to the polar orbit of this planet, we have to pay attention to the tidal factor $f$ from Eq.~\ref{eq.Pdot} in Table~\ref{tab.tidalfactor}, when determining $Q_*'$. The stellar rotation period of KELT-9 is $P_* = (18.96\pm0.34)$\,h \citep{jones2022}, with the orbital period of KELT-9\,b being nearly double the rotation period of the star, with a value of around $P=1.48\,$d \citep{2022AJ....163..122C}. 
This leads to a smaller absolute value of the tidal factor of $f=-135/16$ instead of $-27/2$.
With this, we can also give a $95\%$ confidence lower limit on $Q_*'$ of $Q_{*}' > (9.0\pm3.7)\times 10^4$, with the best-fit parameter being $Q'_{*, \mathrm{best}} = (1.7\pm0.7)\times 10^5$.
This would place KELT-9 near the lower edge of the theoretical predictions, which range from $10^5$ to $10^{8.5}$ in the literature \citep{Meibom2005, Jackson2008, hansen2010, husnoo2012, penev2012, bonomo2017, penev2018, Patel2022}, if the decay trend would be true. Some of these studies assumed $Q_*$ to be a universal constant, however. Still, this value is on a similar level to the tidal quality factor of WASP-12 with $Q_*'=1.75\times10^5$.\\

Nevertheless, since the apsidal precession model gives the best fit for now and we get a non-zero eccentricity ($e = 0.00122$) from it, we can compute the expected tidal circularization timescale $\tau_e = \frac{e}{|\dot{e}|}$.
Using Eq.~\ref{eq.edot} with $f_e=\frac{83}{16}$ from Table~\ref{tab.tidalfactore} with the corresponding system parameters for KELT-9, Jupiter's value for the second Love number ($k_2 = 0.565$) from the Juno mission \citep{2020GeoRL..4786572D}, an estimate for its tidal quality factor ($Q_p = 5\times 10^5$, e.g. \citet{1966Icar....5..375G, 2005ApJ...635..688W}), and the stellar modified quality factor from our tidal decay analysis, yields
\begin{equation}
    \tau_e = 0.78\,\mathrm{Myr}. 
\end{equation}
This estimate indicates that the circularization of the orbit should be completed before the planet would be tidally disrupted by the star, should orbital decay also be happening. 
This is especially the case considering that the semi-major axis would be shrinking during that time as well, leading to an acceleration of the circularization process. 
Concerning the origin of the eccentricity, it is possible that it is justified by the migration history of the planet, for example if the planet migrated inwards due to high-eccentricity migration or planet-planet scattering. Depending on the final eccentricity after the migration, it is plausible that the eccentricity could be maintained until now, especially considering the relatively small value which the apsidal precession fit provides. Another possibility is that the eccentricity could be excited by an unseen third body in the system.

\subsubsection{KELT-16\,b}
\subsubsection*{Transit fitting}
\begin{figure}[h]
    \centering
    \includegraphics[width=\linewidth]{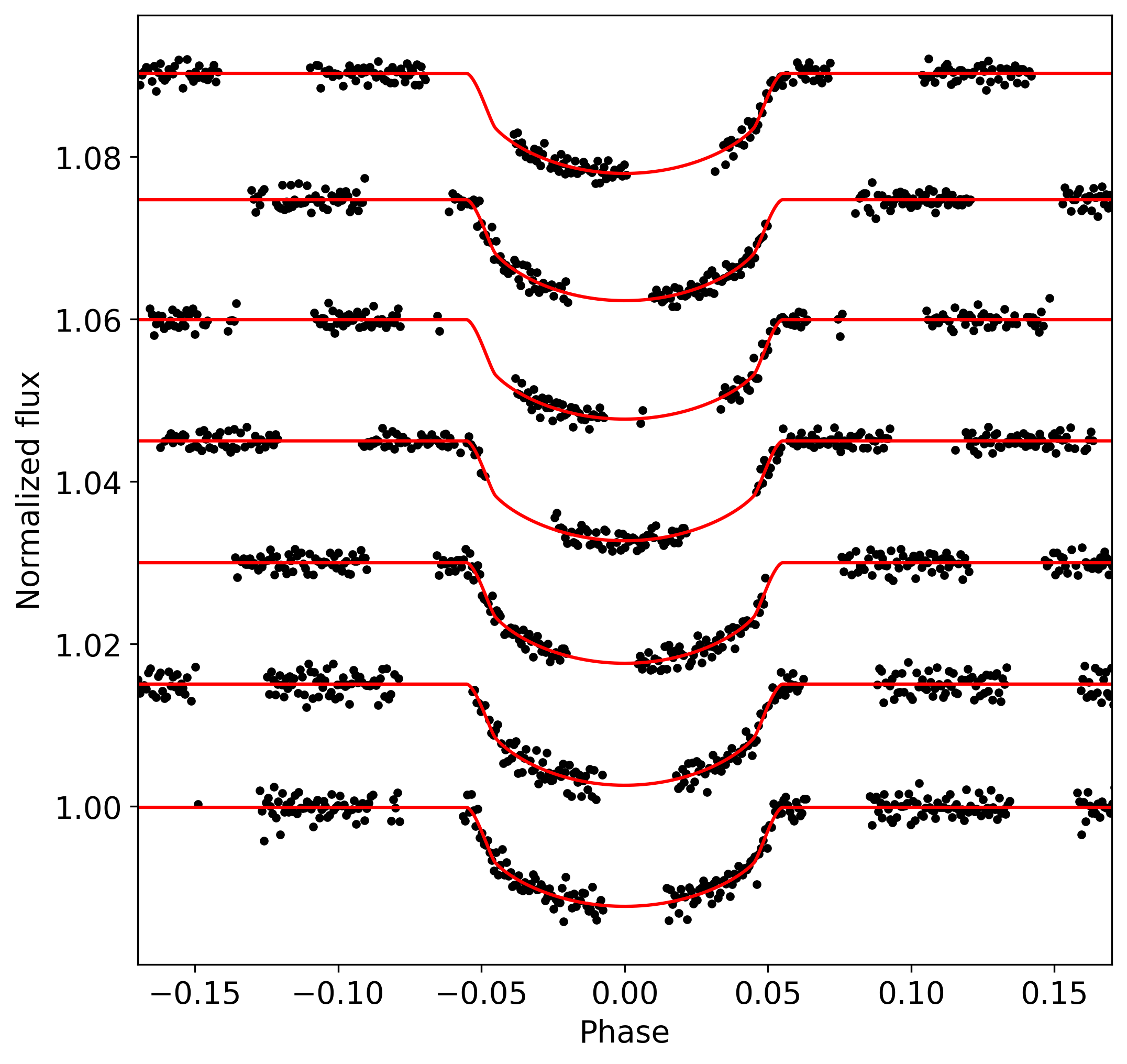}
    \caption{KELT-16\,b transits observed with CHEOPS. The x-axis shows the orbital phase of the transits and the y-axis shows the normalized flux. There is an offset of 0.015 in terms of normalized flux between the individual transits. The data points (black dots) were corrected for the roll angle of the satellite using \textsc{pycheops} and fitted with \textsc{tlcm}. The resulting transit models are indicated by the red lines.}
    \label{fig:KELT-16_transits}
\end{figure}

For KELT-16\,b, we fitted a total of 151 transits. All of these, except those of CHEOPS, were modelled individually with \textsc{tlcm}, since many different telescopes and filters were used for the ground-based observations. 
Even though there are many transit observations, they offer less precision in comparison to KELT-9\,b. The CHEOPS light curves suffer from gaps in between observations, especially if they are located near the ingress or egress phase of the transit (see Fig.~\ref{fig:KELT-16_transits}), meaning the resulting mid-transit times cannot be as precise as the data would allow \citep{barros2013, Borsato2021}.
Because of this, a combined model was created and applied to the final transit fitting of all seven CHEOPS light curves.
The previously unpublished observed transits from CHEOPS are shown in Fig.~\ref{fig:KELT-16_transits} and the obtained mid-transit times can be found in Table~\ref{tab:KELT-16_tts}.

\begin{table*}[h]
    \centering
    \setlength{\tabcolsep}{10pt} 
    \renewcommand{\arraystretch}{1.05} 
    \caption{KELT-16\,b mid-transit times and errors.}
    \begin{tabular}{c c c c}
        \hline
        Mid-transit time & Error & Epoch & Source\\
        $[\mathrm{BJD}_\mathrm{TDB}]$ & [d] & & \\\hline
        2457165.85120 &	0.00076 & -1177 &	\citet{2017AJ....153...97O} \\
        2457166.82162 &	0.00076 & -1176 &	\citet{2017AJ....153...97O} \\
        2457166.82490 &	0.00114 & -1176 &	\citet{2017AJ....153...97O} \\
        2457168.75747 &	0.00132 & -1174 &	\citet{2017AJ....153...97O} \\
        2457196.86000 &	0.00132 & -1145 &    \citet{2017AJ....153...97O} \\
        \dots & \dots & \dots & \dots\\
        \hline
    \end{tabular}
    \tablefoot{Epochs are relative to the middle of the data set. The ``Source'' column describes the source of the respective light curve. Only a portion of the data is shown here, the full table will be made available with the online version of this paper or at the CDS.}
    \label{tab:KELT-16_tts}
\end{table*}

\subsubsection*{Timing analysis}
\begin{figure}[h]
    \centering
    \includegraphics[width=\linewidth]{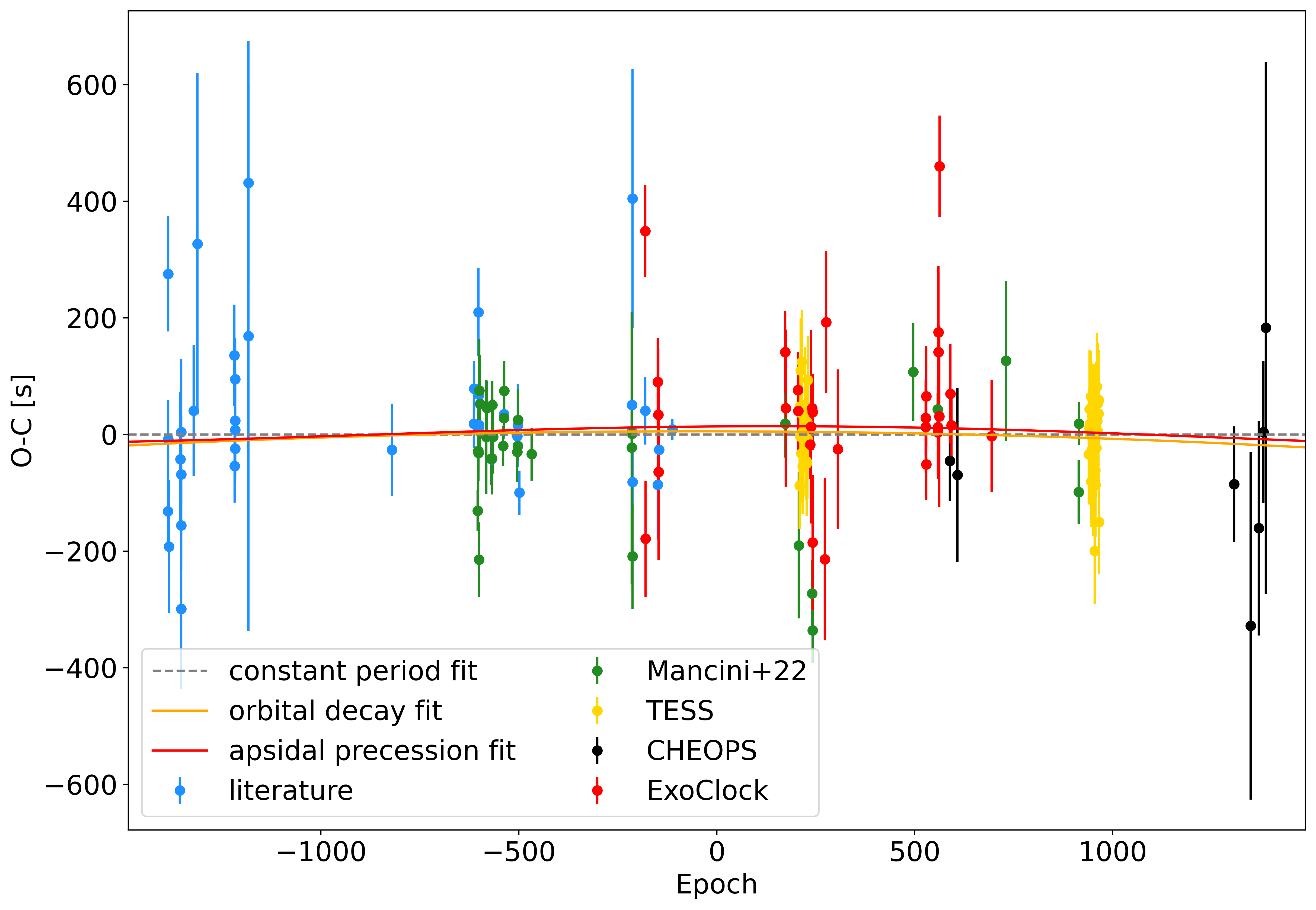}
    \caption{O-C plot showing the deviations in transit time from the best-fit linear ephemeris (gray dashed line) for KELT-16\,b. The transit number is shown on the x-axis, and the timing difference on the y-axis. The orange line shows the best quadratic fit to the KELT-16\,b transit timing variation data, with the red line showing the best apsidal precession fit. CHEOPS data are highlighted in black. The data from the literature is described in more detail in section~\ref{sec:obs_KELT-16}.}
    \label{fig:Kelt-16_ttvs}
\end{figure}

Fitting the obtained times of mid-transit using the three models, as described in Section~\ref{sect:fit_descr}, we obtain Fig. \ref{fig:Kelt-16_ttvs}. 
The fit parameters of the three models can be found in Table~\ref{tab:fit_parameters}.\\

Since we do not have observations of secondary eclipses, and those observed by TESS suffer from a low signal-to-noise ratio due to the faintness of the target, we cannot obtain accurate mid-occultation times. This limits us in examining the difference of tidal orbital decay and apsidal precession. However, the Keplerian orbit model is the favoured model anyway, having the lowest BIC value at $BIC_k = 292.9$, with the orbital decay model leading to $BIC_d = 297.6$, and the apsidal precession model yielding $BIC_p = 311.0$. As with KELT-9, the orbital periods resulting from the fits agree with each other well within $1\,\sigma$. Besides that, the $dP/dN$ of the orbital decay model also agrees with a constant period within $1\,\sigma$, with the resulting eccentricity $e$ of the apsidal precession model agreeing with a circular orbit at $2\,\sigma$. The $d\omega/dN$ parameter agrees with a Keplerian orbit at only just over $4\,\sigma$.
Comparing our orbital decay fit results to earlier results from the literature, we find that our result of $dP/dN = (-2.73 \pm 3.77)\times10^{-10}\,$d/orbit from our independent analysis agrees well with the values of $dP/dN=(-0.1\pm1.4)\times10^{-9}\,$d/orbit from \citet{2018AcA....68..371M}, $dP/dN = (-0.6\pm1.4)\times 10^{-9}\,$d/orbit from \citet{2020AJ....159..150P}, and $dP/dN = (-3.2\pm4.0)\times 10^{-10}\,$d/orbit from \citet{2022MNRAS.509.1447M}. These literature results also all agree with a constant orbital period.
In general, it can be said that for this object there is an overall lack of precision to find an effect that is as small as the one we are looking for, leading to scatter in the data and relatively large error bars. A longer baseline of observations is needed, with the current baseline only spanning about $6.5\,$yr. However, this will be extended with TESS in Sector 55, where KELT-16 will be located in the field-of-view of camera 2. Nevertheless, we can use our tidal decay fit result to constrain the modified stellar tidal quality factor, assuming that the tidal decay trend is true. Our calculated lower limit of $Q_*' > (2.1\pm 0.9)\times 10^5$ for this is close to the one of KELT-9, and with that also close to the actual $Q_*'$ value of WASP-12, meaning that we expect a slower orbital decay for KELT-16\,b than for WASP-12b. Moreover, our lower limit value is in agreement with the one from \citet{2022MNRAS.509.1447M} with $(1.9\pm0.8)\times 10^5$. Using our best-fit value of $\dot{P} = (-8.94 \pm 12.35)\,$ms\,yr$^{-1}$ to calculate the orbital decay timescale $\tau$, we obtain a $95\%$ confidence lower limit of $\tau > (2.5\pm 0.9)\,\mathrm{Myr}$.

\subsubsection{WASP-4\,b}
\subsubsection*{Transit fitting}
\begin{figure}[h]
    \centering
    \includegraphics[width=\linewidth]{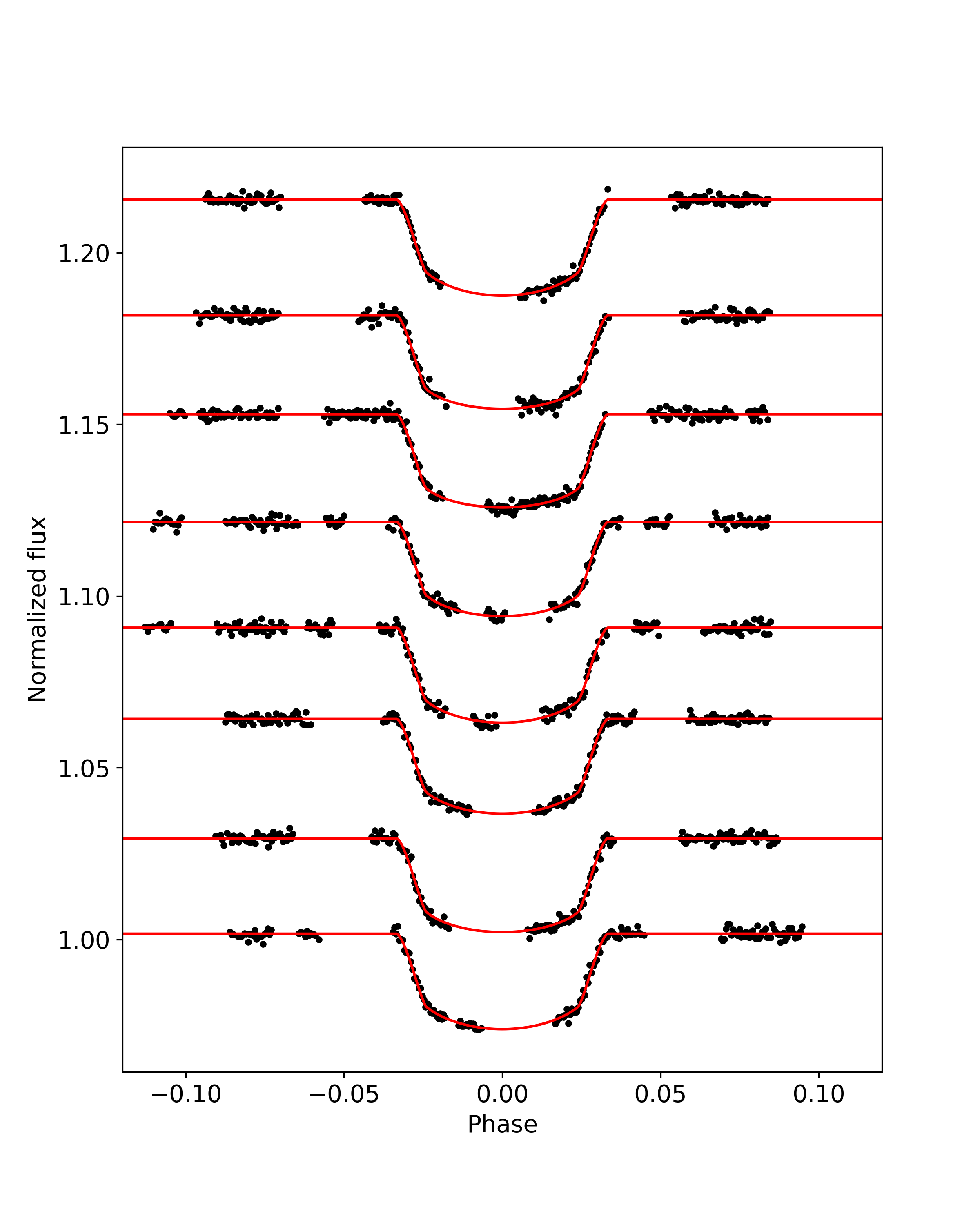}
    \caption{Transits of WASP-4\,b observed with CHEOPS. The x-axis shows the orbital phase and the y-axis the normalized flux. There is an offset of 0.03 in terms of normalized flux between the individual transits. The data points (black dots) were corrected for the roll angle of the satellite using \textsc{pycheops} and fitted with \textsc{tlcm}.}
    \label{fig:WASP-4_transits}
\end{figure}

We have a total of 172 transit timing data points for this target, many of which offer high precision. To test the validity of the timings obtained by \citet{2020MNRAS.496L..11B}, we compared them against those from the analysis of \citet{2019MNRAS.490.4230S}, their own earlier paper \citep{2019MNRAS.490.1294B}, and us. Particularly, the timings with very high precision were re-analysed and verified. Still, there is the possibility that the errors on the measured flux may be underestimated in some of those cases. However, they were recorded with high precision ground-based instruments, like those of the GEMINI observatory or ESO's VLT.
All but six of their 29 timings from the light curves of \citet{2009AA...496..259G}, \citet{2011ApJ...733..127S} and \citet{2019MNRAS.490.4230S} agree with our results within $3\,\sigma$, with a median value of $0.69\,\sigma$.
The ExoClock data are scattered and in some cases still only have relatively small error bars, which could be caused by underestimated uncertainties in the original light curves.\\
The transits observed with CHEOPS for this target are shown in Fig.~\ref{fig:WASP-4_transits}, and the literature and newly fitted transit times can be found in Table~\ref{tab:WASP-4_tts}.

\begin{table*}[h]
    \centering
    \setlength{\tabcolsep}{10pt} 
    \renewcommand{\arraystretch}{1.05} 
    \caption{WASP-4\,b mid-transit times and errors. }
    \begin{tabular}{c c c c}
        \hline
        Mid-transit time [BJD$_\mathrm{TDB}$] & Error [d] & Epoch & Source\\
        $[\mathrm{BJD}_\mathrm{TDB}]$ & [d] & & \\\hline
        2453960.43148 & 0.00182 & -2071 & WASP \\
        2454361.90048 & 0.00188 & -1771 & WASP \\
        2454396.69616 & 0.00008 & -1745 & \citet{2020MNRAS.496L..11B} \\ 
        2454697.79815 & 0.00006 & -1520 & \citet{2020MNRAS.496L..11B} \\ 
        2454697.79830 & 0.00013 & -1520 & \citet{2020MNRAS.496L..11B} \\
        \dots & \dots & \dots & \dots\\
        \hline
    \end{tabular}
    \tablefoot{The epochs are given in relation to the middle of the data set. The ``Source'' column describes the source of the respective light curve or the source of the timing, which is described in more detail in section~\ref{sec:obs}. Only a portion of the data is shown here, the full table will be made available with the online version of this paper and at the CDS.}
    \label{tab:WASP-4_tts}
\end{table*}

\subsubsection*{Timing analysis}
\begin{figure}[h]
    \centering
    \includegraphics[width=\linewidth]{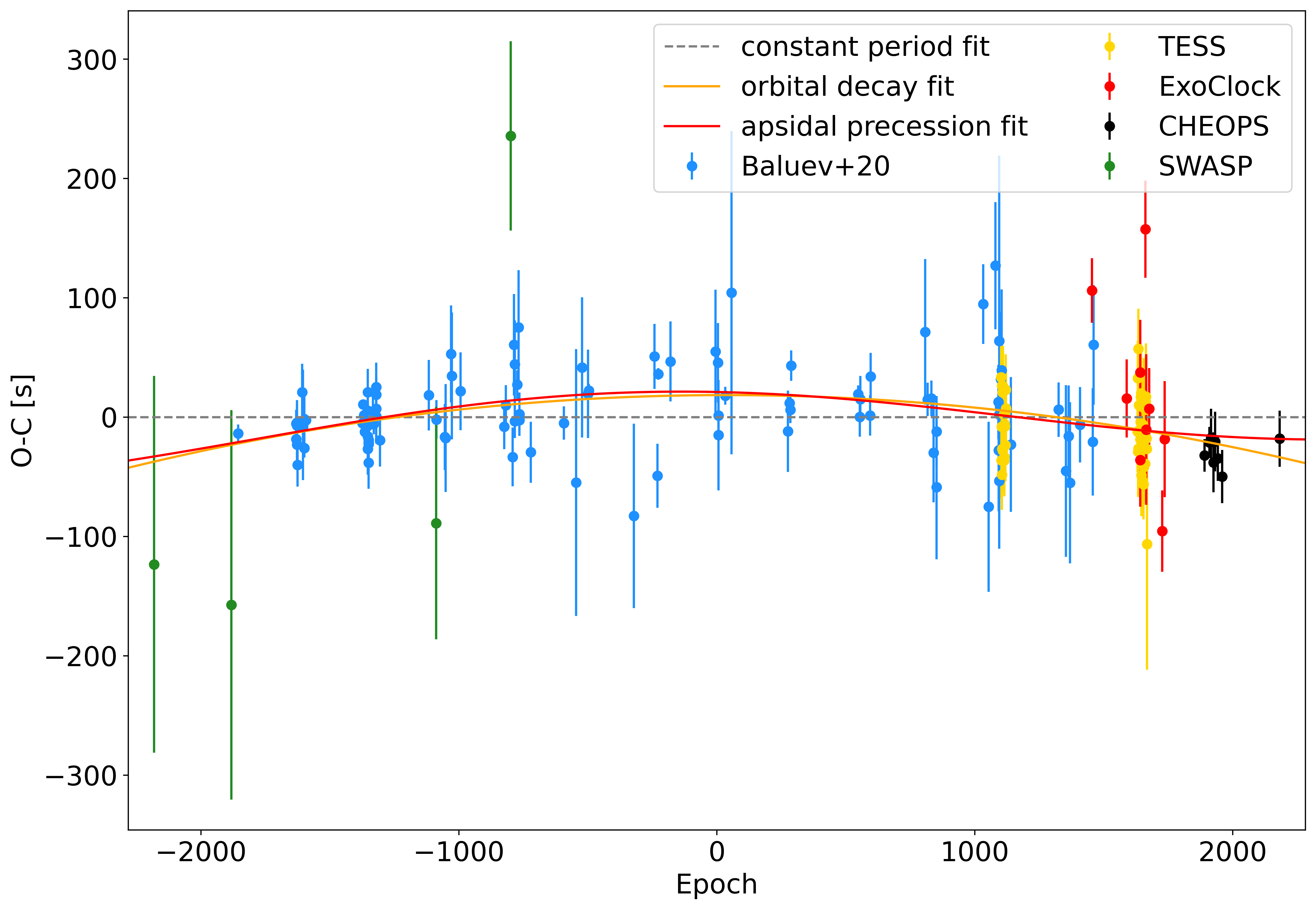}
    \caption{O-C plot showing the deviations in transit time from the best-fit linear ephemeris (gray dashed line) for WASP-4\,b. The transit number is shown on the x-axis, with the difference in timing being shown on the y-axis. The orange line shows the quadratic fit to the WASP-4\,b transit timing variation data, and the red line shows the apsidal precession fit. CHEOPS data are highlighted in black.}
    \label{fig:WASP-4_ttvs}
\end{figure}

\begin{figure}[h]
    \centering
    \includegraphics[width=\linewidth]{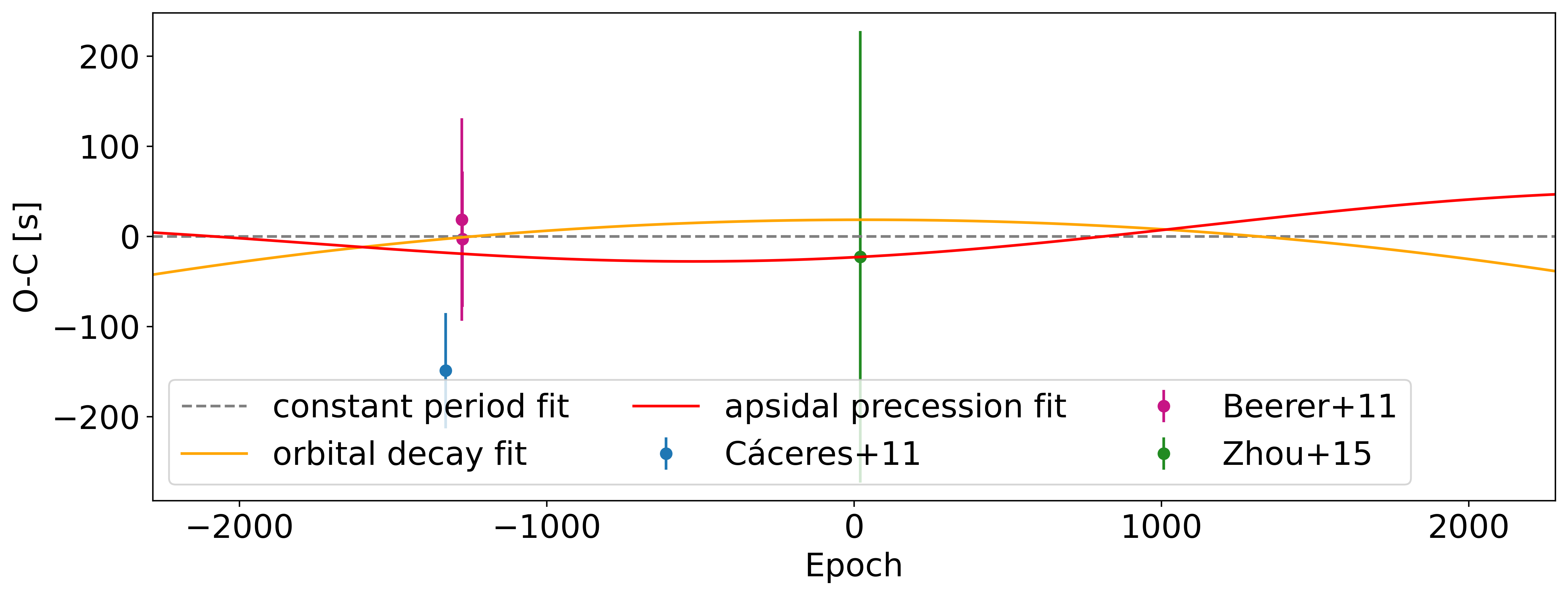}
    \caption{Occultation timings for WASP-4\,b from the literature, with the three lines representing our three models according to the legend. The x-axis shows the epoch relative to the middle of the data set, and the y-axis shows the difference in measured and calculated (linear) timing.}
    \label{fig:WASP-4_occs}
\end{figure}

Fitting the obtained timings as described in Section~\ref{sect:fit_descr} using the MCMC algorithms
, yields the models and parameters in Fig.~\ref{fig:WASP-4_ttvs}, Fig.~\ref{fig:WASP-4_occs} and in Table~\ref{tab:fit_parameters}.

From the final fit parameters, we can see that the orbital decay model is the preferred model with the lowest BIC value of $BIC_d = 244.63$. The apsidal precession and Keplerian orbit models lead to higher BIC values of $BIC_p = 252.31$ and $BIC_k = 339.52$, respectively. There is no clear favoured model from the occultation measurements due to the scatter of the data and their error bars. There are also only four secondary eclipse timings in the literature and the eclipses are not visible with TESS, as \citet{2021arXiv211209621T} also found. This hinders differentiation between the models using only the available occultations.\\

For this target, the best-fit orbital period values agree with each other within $1\,\sigma$.
The best-fit $dP/dN$ value of the orbital decay model is non-zero at a $5\,\sigma$ level, disagreeing with a linear ephemeris and making this trend significant.
Nevertheless, the eccentricity value of the apsidal precession model is also in disagreement with a Keplerian orbit at more than $3\,\sigma$.
This, in combination 
with the position of the CHEOPS data points in the O-C plot in Fig.~\ref{fig:WASP-4_ttvs}, indicates that the system is not Keplerian.
More observations in the future are needed to help to differentiate between the two preferred models, and TESS could contribute to this with a planned observation of this target during Sector 69 of its second extended mission in camera 2 from August to September 2023.\\
Assuming that the fitted orbital decay trend is true, we find that our derived decay rate of $dP/dN=(-2.62\pm0.49)\times10^{-10}\,$d/orbit agrees with the values previously given in the literature, which range from $(-1.0\pm1.0)\times10^{10}\,$d/orbit in the study of \citet{2019MNRAS.490.1294B}, which was adjusted to $(-1.7\pm0.5)\times10^{-10}\,$d/orbit after re-assessing older data and adding new data \citep{2020MNRAS.496L..11B}, to $(-5.4\pm0.5)\times10^{-10}\,$d/orbit \citep{2019AJ....157..217B}, with other literature values lying between them \citep{2019MNRAS.490.4230S, 2020ApJ...893L..29B, 2021arXiv211209621T}. The most recent value from \citet{2022arXiv220211990M} with $dP/dN = (-2.1\pm0.6)\times10^{-10}$ is in good agreement with our analysis.
Under the same assumption, we can also compare our lower limit on $Q_*'$ of $Q_{*}' > (4.1\pm0.9)\times10^4$ and best-fit value of $Q_{*,\mathrm{best}}' = (5.7 \pm 1.0)\times 10^{4}$ with the literature values and notice that it agrees with the above mentioned publications, and
that it is significantly lower than the theoretical estimates stated in Section~\ref{sec:KELT-9_disc}. If we calculate the potential orbital decay timescale of WASP-4\,b, we get:
\begin{equation*}
    \tau = 18.73 \pm 3.48 \,\mathrm{Myr,}
\end{equation*}
which leads to a $95\%$ confidence lower limit of $\tau_\mathrm{low} = (13.7\pm 1.8)\,$Myr.
Comparing this to the WASP-12 system, we notice that the orbital decay timescale is much shorter there with $\tau = 3.25^{+0.24}_{-0.21}$\,Myr, even though the tidal quality factor of WASP-12 is higher with $1.75^{+0.13}_{-0.11}\times10^5$ as compared to WASP-4 with $(5.6\pm1.0)\times10^4$, if we assume that orbital decay is actually happening. However, this can be explained by the combination of the system parameters of these two systems, namely $M_P$, $M_*$, $R_*$ and $a$, and their influence in Eq.~\ref{eq:qstar}.

\subsubsection{HD\,97658\,b}

\subsubsection*{Transit fitting}
\begin{figure}[h]
    \centering
    \includegraphics[width=\linewidth]{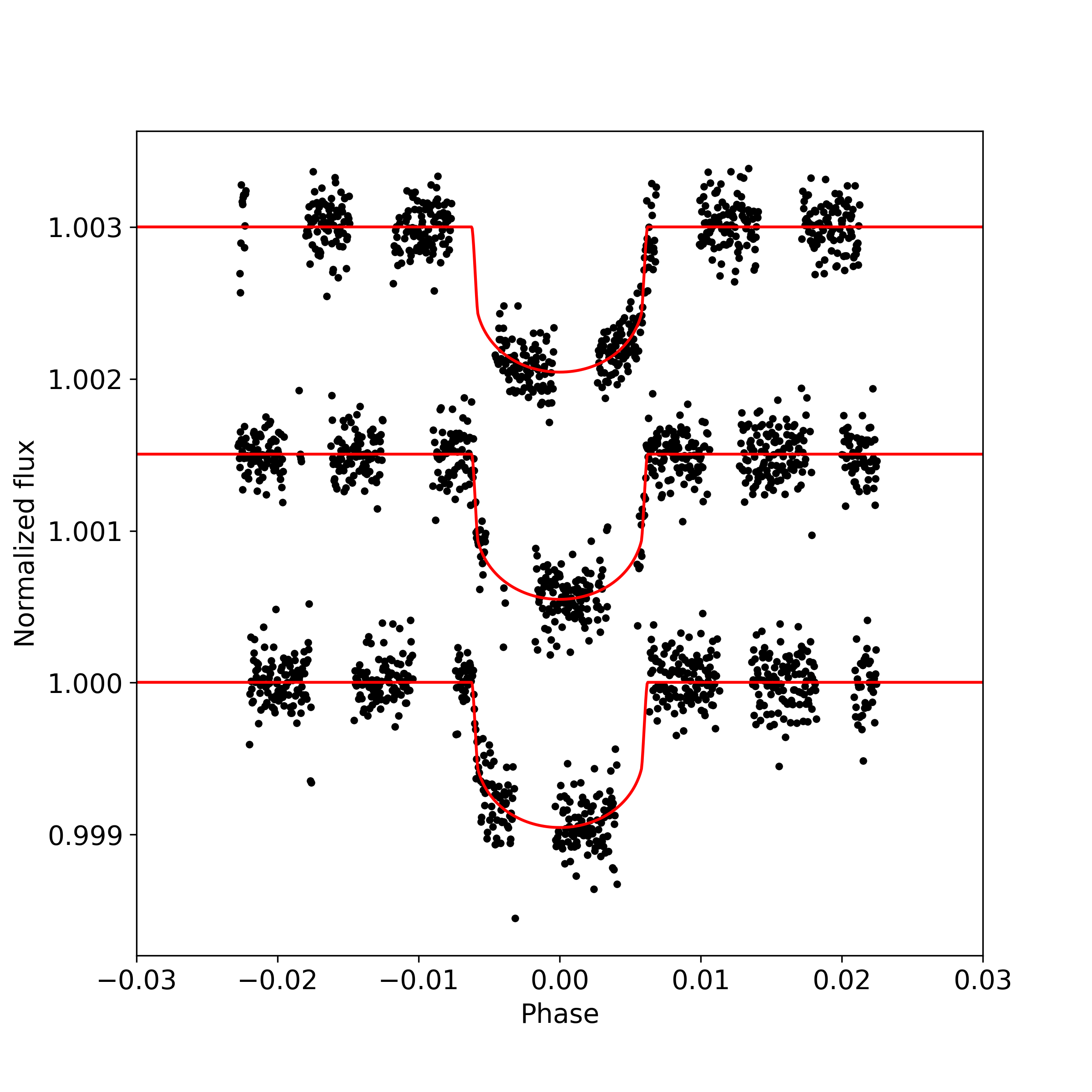}
    \caption{New transit observations of HD\,97658\,b with CHEOPS. The x-axis shows the orbital phase, and the y-axis shows the normalized flux. There is an offset of 0.0015 in terms of normalized flux between the individual transits. The data points (black dots) were corrected for the roll angle of the satellite using \textsc{pycheops} and fitted with \textsc{tlcm}. The resulting transit models are shown with red lines for each transit.}
    \label{fig:HD97658_transits}
\end{figure}

We fitted the four available CHEOPS transits using \textsc{tlcm} by first creating a combined model and then fitting the transits individually using the parameters from the combined model, which is useful due to the gaps in the light curve. For the TESS Sector 49 data, we fitted the data of the whole sector together using the SAP light curve, to get a more precise timing than what would be possible using individual fits, and because it was already done this way for the earlier TESS observation. The SAP light curve was chosen over the PDCSAP light curve because the latter excludes one of the transits near the center of the light curve, and the last transit is on a slope, which it is not in the SAP light curve. 
The SAP light curve was reasonably flat in each transit cut-out and thus required no further detrending.
Nevertheless, we still fitted the transits using the PDCSAP flux as well, which resulted in similar mid-transit times ($< 0.12\,\sigma$ difference) and also error bars in comparison to the individual transit fits using the SAP flux.
The remaining mid-transit times were taken from \citet{2021MNRAS.tmp.3057M} to have a direct comparison. The newly observed CHEOPS transits are shown in Fig.~\ref{fig:HD97658_transits}, the fit to the new TESS observation in Fig.~\ref{fig:HD97658_tess_transits}, and all obtained mid-transit times can be found in Table~\ref{tab:HD97658_tts}.

\subsubsection*{Timing analysis}
\begin{figure}[h]
    \centering
    \includegraphics[width=\linewidth]{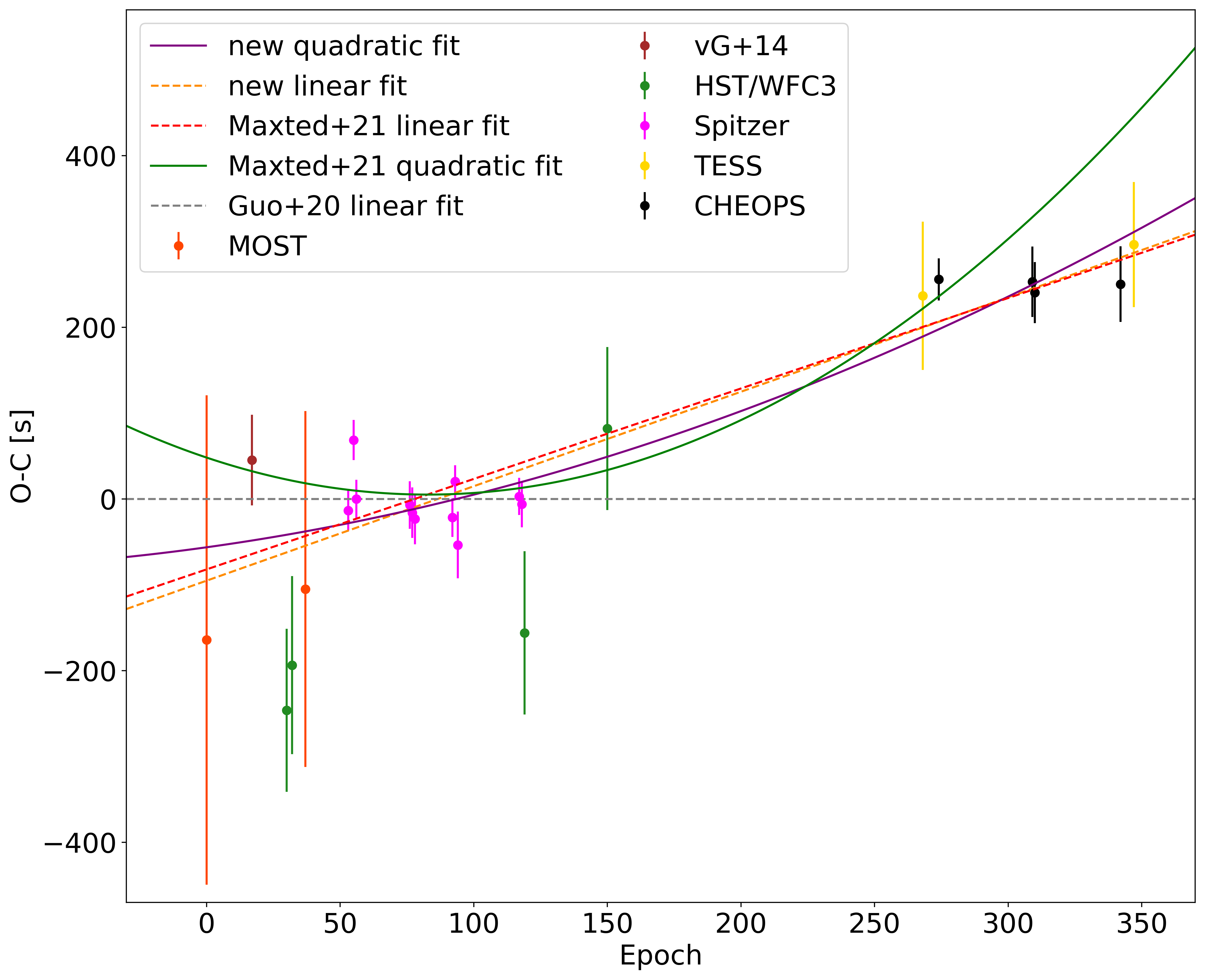}
    \caption{O-C plot showing the deviations in transit time from the best-fit linear ephemeris from \citet{2020AJ....159..239G} (gray dashed line) for HD\,97658\,b to have a direct comparison to the \citet{2021MNRAS.tmp.3057M} best-fit linear and quadratic models (red dashed and green solid lines). Our best-fit models are shown as an orange dashed line for the linear model, and as a purple solid line for the quadratic model. The transit number is shown on the x-axis, and the difference in timing is shown on the y-axis. CHEOPS data are highlighted in black.}
    \label{fig:HD97658_ttvs}
\end{figure}

The timing analysis for this target was done as described in Section~\ref{sect:fit_descr}, with the exception of an apsidal precession fit. 
Besides the new quadratic and linear fits for this target, we also show the fits of \citet{2020AJ....159..239G} and \citet{2021MNRAS.tmp.3057M} in Fig.~\ref{fig:HD97658_ttvs}. Our fit parameters can be found in Table~\ref{tab:fit_parameters}.

Comparing our best-fit results for a linear and a quadratic model to the now available data with the linear fit of \citet{2020AJ....159..239G} and the linear and quadratic fits of \citet{2021MNRAS.tmp.3057M}, we find that the orbital period found in the former publication of $P_\mathrm{G} = (9.489295\pm0.000005)\,$d still agrees with our linear best-fit period of $P = (9.48930773\pm0.00000149)\,$d at just over $3\,\sigma$. The updated orbital period of \citet{2021MNRAS.tmp.3057M} with a value of $P_\mathrm{M} = (9.4893072\pm0.0000025)\,$d improves the earlier result and agrees with our findings at well within $1\,\sigma$.
Comparing the quadratic fits, which was their preferred model by means of BIC, they find an orbital period of $P_\mathrm{M,q} = (9.4892968\pm0.0000038)\,$d, and period change per orbit of $dP/dN_\mathrm{M} = (1.46\pm0.48)\times10^{-7}\,$d/orbit, whereas we find $P_\mathrm{q} = (9.48929994\pm0.00000824)\,$d and $dP/dN = (0.42\pm0.44)\times 10^{-7}$ using the now available data. 
The quadratic model shows weak evidence against the linear model with $\Delta \mathrm{BIC} = 58.13 - 55.05 = 3.08$, but also agrees with a linear trend within $1\,\sigma$, rendering this as inconclusive. The scatter in the early data contributes to this significantly. 
Looking at the Spitzer transits in Fig.~9 in the publication of \citet{2020AJ....159..239G}, one can notice the relatively shallow transit depth of the transits in relation to the noise of the data. Comparing this Figure with the resulting uncertainties on the transit timings might indicate that the uncertainties of the transit fits could be underestimated. This could explain the scatter of this data set in relation to their error bars and our best-fit models. This suspicion is supported by the fact that \citet{2014ApJ...786....2V} also analysed one Spitzer transit recorded with Spitzer's IRAC at $4.5\mu$m, with their analysis yielding approximately double the uncertainty on the timing. However, improvements of the Spitzer data reduction pipeline or improved fitting methods, might also lead to more precise transit fits. An important thing to mention is that there are only few outliers though, and that most of the Spitzer mid-transit times analysed by \citet{2020AJ....159..239G} are in agreement with both of the linear models.\\
Concluding, by taking new observations of the transits of HD\,97658\,b, we find a weaker quadratic trend than \citet{2021MNRAS.tmp.3057M}, and so are unable to conclusively distinguish between a linear and a quadratic trend.


\begin{table*}[h]
    \centering
    \setlength{\tabcolsep}{10pt} 
    \renewcommand{\arraystretch}{1.2} 
    \caption{Fit results of all models for KELT-9\,b, KELT-16\,b, WASP-4\,b, and HD\,97658\,b. }
    \begin{tabular}{c c c c}
        \hline
        Planet & Model & Parameter & Value \\\hline\hline
        KELT-9\,b & Constant period     & $t_0$ $[\mathrm{BJD}_\mathrm{TDB}]$ & $2458162.09128 \pm 0.00009$ \\
                            && $P$ [d]       & $1.48111916 \pm 0.00000013$ \\
                            && BIC           & 103.73 \\\cline{2-4}
        & Orbital decay       & $t_0$ $[\mathrm{BJD}_\mathrm{TDB}]$        & $2458162.09150 \pm 0.00013$ \\
                            && $P$ [d]        & $1.48111926 \pm 0.00000013$   \\
                            && $dP/dN$ [d/orbit]& $(-11.41\pm4.98)\times 10^{-10}$   \\
                            && BIC           & 95.28         \\\cline{2-4}
        & Apsidal precession  & $t_0$ $[\mathrm{BJD}_\mathrm{TDB}]$         & $2458162.09119 \pm 0.00009$ \\
                            && $P_s$ [d]     & $1.48111973 \pm 0.00000017$   \\
                            && $e$           & $0.00122 \pm 0.00025$   \\
                            && $\omega_0$ [rad]  & $4.426\pm0.254$   \\
                            && $d\omega/dN$ [rad/orbit] & $0.001624\pm0.000300$   \\
                            && BIC           & 90.45 \\\hline
        KELT-16\,b & Constant period     & $t_0$ $[\mathrm{BJD}_\mathrm{TDB}]$ & $2458509.84580\pm 0.00010$\\
                            && $P$           & $0.96899284\pm0.00000014$ \\
                            && BIC           & 292.90 \\\cline{2-4}
        & Orbital decay       & $t_0$ $[\mathrm{BJD}_\mathrm{TDB}]$ & $2458509.84586\pm0.00013$ \\
                            && $P$ [d]       & $0.96899283\pm0.00000014$   \\
                            && $dP/dN$ [d/orbit] & $(-2.73\pm3.77)\times 10^{-10}$ \\
                            && BIC           & 297.63 \\\cline{2-4}
        & Apsidal precession  & $t_0$ $[\mathrm{BJD}_\mathrm{TDB}]$         & $2458509.84576\pm0.00013$   \\
                            && $P_s$ [d]     & $0.96899281\pm0.00000023$   \\
                            && $e$           & $0.00065\pm0.00047$   \\
                            && $\omega_0$ [rad] & $2.838\pm1.685$   \\
                            && $d\omega/dN$ [rad/orbit] & $0.001441\pm0.000335$   \\
                            && BIC           & 310.95           \\\hline
        WASP-4\,b & Constant period     & $t_0$ $[\mathrm{BJD}_\mathrm{TDB}]$         & $2456880.45366\pm0.00003$   \\
                            && $P$ [d]       & $1.33823132\pm0.00000002$   \\
                            && BIC           & 351.03           \\\cline{2-4}
        & Orbital decay       & $t_0$ $[\mathrm{BJD}_\mathrm{TDB}]$         & $2456880.45387\pm0.00005$   \\
                            && $P$ [d]       & $1.33823133\pm0.00000002$   \\
                            && $dP/dN$ [d/orbit] & $(-2.62\pm0.49)\times 10^{-10}$   \\
                            && BIC           & 245.34         \\\cline{2-4}
        & Apsidal precession  & $t_0$ $[\mathrm{BJD}_\mathrm{TDB}]$         & $2456880.45359\pm0.00005$   \\
                            && $P_s$ [d]     & $1.33823140\pm0.00000008$   \\
                            && $e$           & $0.00079\pm0.00020$   \\
                            && $\omega_0$ [rad] & $3.492\pm0.310$   \\
                            && $d\omega/dN$ [rad/orbit] & $0.001089\pm0.000103$   \\
                            && BIC           & 252.25           \\\hline
        HD\,97658\,b & Constant period     & $t_0$ $[\mathrm{BJD}_\mathrm{TDB}]$         & $2456361.80580\pm0.00024$   \\
                            && $P$ [d]       & $9.48930773\pm0.00000149$   \\
                            && BIC           & 58.13           \\\cline{2-4}
        & Quadratic fit       & $t_0$ $[\mathrm{BJD}_\mathrm{TDB}]$         & $2456361.80625\pm0.00053$   \\
                            && $P$ [d]       & $9.48929994\pm0.00000824$   \\
                            && $dP/dN$ [d/orbit] & $(4.20\pm4.37)\times 10^{-8}$   \\
                            && BIC           & 55.05           \\
        \hline
    \end{tabular}
    \tablefoot{BIC stands for Bayesian Information Criterion, the lower the value, the better the fit, but the number of parameters is also taken into account.}
    \label{tab:fit_parameters}
\end{table*}




\section{Conclusions\label{sec:conclusion}}

We homogeneously analysed new CHEOPS and TESS photometric data, as well as re-analysed archival photometric data for KELT-9\,b, KELT-16\,b, WASP-4\,b, and HD\,97658\,b, using state-of-the-art software, like \textsc{tlcm} and \textsc{pycheops}, and developed own MCMC routines for the analysis of this data. Transit fitting was done using \textsc{tlcm}, occultation fitting using \textsc{batman} in combination with an MCMC approach, and the transit timing variations for each system were fitted using MCMC algorithms for three different models. The first of these models is a constant period model, assuming a Keplerian orbit, the second is an orbital decay model, accounting for a changing orbital period, and the third is an apsidal precession model, assuming that the orbit is slightly eccentric, leading to a precessing orbit which can mimic orbital decay for short baselines, but can be distinguished using secondary eclipses or having long baselines.\\

For the targets KELT-9\,b and KELT-16\,b we did a homogeneous re-analysis of every available transit from the literature and added new observations to these.
For KELT-9\,b, we find that the timing deviations are best described by an apsidal precession model with an eccentricity of $e = 0.00122 \pm 0.00025$, which could, however, already be different with new observations, for example those of TESS. Yet, in general, apsidal precession does not rule out tidal decay and vice versa, a combination of both is likely.
The KELT-16 system is best described with a Keplerian orbit model since the scatter and error bars are both relatively large for the observations of this system. Nevertheless, there will also soon be new TESS observations of this system available which could give an indication towards one of the models.\\
For WASP-4\,b, we re-analysed data from a recent publication \citep{2021arXiv211209621T}, as well as TESS data and added new CHEOPS transit observations to the data set. The remaining data were taken from a recent study which also re-analysed earlier transit data in a homogeneous way \citep{2020MNRAS.496L..11B}.
We find the orbital decay model to describe the data the best and find a similar significant trend at $5\,\sigma$ as earlier evaluations of this system \citep{2020ApJ...893L..29B, 2020MNRAS.496L..11B, 2021arXiv211209621T, ivshina2022}, but apsidal precession cannot be ruled out yet. The TESS observations of this target towards the end of 2023 could already rule out one of the two preferred models.\\
In the case of HD\,97658\,b, we re-analyse the only available CHEOPS transit at the time of the study \citep{2021MNRAS.tmp.3057M} and combine the analysis with three recently acquired transits of this target from CHEOPS. 
Moreover, we also analyse new TESS data from Sector 49, including three transits of this target. 
Adding these new transit timings to the data set of \citet{2021MNRAS.tmp.3057M}, we find a weaker quadratic trend than them, which only deviates from a linear trend at $1\,\sigma$\, leaving this as inconclusive.


\begin{acknowledgements}
We thank the anonymous referee for their helpful comments.
The CHEOPS photometric data underlying this article is publicly accessible using the CHEOPS archive browser\footnote{\href{https://cheops-archive.astro.unige.ch/archive_browser/}{https://cheops-archive.astro.unige.ch/archive\_browser/}}, at the DACE website in the CHEOPS database, or via \textsc{pycheops} using the file keys provided in Table~\ref{tab:obs_log}. The TESS data used in this publication is publicly available at the MAST archive. The timing data for the various planets and the photometric CHEOPS data is available with the online version of this article and at the CDS.
    CHEOPS is an ESA mission in partnership with Switzerland with important contributions to the payload and the ground segment from Austria, Belgium, France, Germany, Hungary, Italy, Portugal, Spain, Sweden, and the United Kingdom. The CHEOPS Consortium would like to gratefully acknowledge the support received by all the agencies, offices, universities, and industries involved. Their flexibility and willingness to explore new approaches were essential to the success of this mission. 
    JVH acknowledges the support of the DFG priority programme SPP 1992 “Exploring the Diversity of Extrasolar Planets (SM 4862-1)”.
    S.C.C.B. acknowledges support from FCT through FCT contracts nr. IF/01312/2014/CP1215/CT0004. 
    This project has received funding from the European Research Council (ERC) under the European Union’s Horizon 2020 research and innovation programme (project {\sc Four Aces}. 
    PM acknowledges support from STFC research grant number ST/M001040/1. 
    LMS gratefully acknowledges financial support from the CRT foundation under Grant No. 2018.2323 ‘Gaseous or rocky? Unveiling the nature of small worlds’.
    B.-O.D. acknowledges support from the Swiss National Science Foundation (PP00P2-190080). 
    This work was granted access to the HPC resources of MesoPSL financed by the Region Ile de France and the project Equip@Meso (reference ANR-10-EQPX-29-01) of the programme Investissements d'Avenir supervised by the Agence Nationale pour la Recherche. 
    YA and MJH acknowledge the support of the Swiss National Fund under grant 200020\_172746. 
    We acknowledge support from the Spanish Ministry of Science and Innovation and the European Regional Development Fund through grants ESP2016-80435-C2-1-R, ESP2016-80435-C2-2-R, PGC2018-098153-B-C33, PGC2018-098153-B-C31, ESP2017-87676-C5-1-R, MDM-2017-0737 Unidad de Excelencia Maria de Maeztu-Centro de Astrobiología (INTA-CSIC), as well as the support of the Generalitat de Catalunya/CERCA programme. The MOC activities have been supported by the ESA contract No. 4000124370. 
    XB, SC, DG, MF and JL acknowledge their role as ESA-appointed CHEOPS science team members. 
    ABr was supported by the SNSA. 
    ACC acknowledges support from STFC consolidated grant numbers ST/R000824/1 and ST/V000861/1, and UKSA grant number ST/R003203/1. 
    This project was supported by the CNES. 
    The Belgian participation to CHEOPS has been supported by the Belgian Federal Science Policy Office (BELSPO) in the framework of the PRODEX Programme, and by the University of Liège through an ARC grant for Concerted Research Actions financed by the Wallonia-Brussels Federation. 
    L.D. is an F.R.S.-FNRS Postdoctoral Researcher. 
    This work was supported by FCT - Fundação para a Ciência e a Tecnologia through national funds and by FEDER through COMPETE2020 - Programa Operacional Competitividade e Internacionalizacão by these grants: UID/FIS/04434/2019, UIDB/04434/2020, UIDP/04434/2020, PTDC/FIS-AST/32113/2017 \& POCI-01-0145-FEDER- 032113, PTDC/FIS-AST/28953/2017 \& POCI-01-0145-FEDER-028953, PTDC/FIS-AST/28987/2017 \& POCI-01-0145-FEDER-028987, O.D.S.D. is supported in the form of work contract (DL 57/2016/CP1364/CT0004) funded by national funds through FCT. 
    MF and CMP gratefully acknowledge the support of the Swedish National Space Agency (DNR 65/19, 174/18). 
    DG gratefully acknowledges financial support from the CRT foundation under Grant No. 2018.2323 ``Gaseousor rocky? Unveiling the nature of small worlds''. 
    M.G. is an F.R.S.-FNRS Senior Research Associate. 
    SH gratefully acknowledges CNES funding through the grant 837319. 
    KGI is the ESA CHEOPS Project Scientist and is responsible for the ESA CHEOPS Guest Observers Programme. She does not participate in, or contribute to, the definition of the Guaranteed Time Programme of the CHEOPS mission through which observations described in this paper have been taken, nor to any aspect of target selection for the programme. 
    ML acknowledges support of the Swiss National Science Foundation under grant number PCEFP2\_194576. 
    LBo, GBr, VNa, IPa, GPi, RRa, GSc, VSi, and TZi acknowledge support from CHEOPS ASI-INAF agreement n. 2019-29-HH.0. 
    This work was also partially supported by a grant from the Simons Foundation (PI Queloz, grant number 327127). 
    IRI acknowledges support from the Spanish Ministry of Science and Innovation and the European Regional Development Fund through grant PGC2018-098153-B- C33, as well as the support of the Generalitat de Catalunya/CERCA programme. 
    S.G.S. acknowledge support from FCT through FCT contract nr. CEECIND/00826/2018 and POPH/FSE (EC). 
    GyMSz acknowledges the support of the Hungarian National Research, Development and Innovation Office (NKFIH) grant K-125015, a a PRODEX Experiment Agreement No. 4000137122, the Lend\"ulet LP2018-7/2021 grant of the Hungarian Academy of Science and the support of the city of Szombathely. 
    V.V.G. is an F.R.S-FNRS Research Associate. 
    NAW acknowledges UKSA grant ST/R004838/1. 
    ACC and TW acknowledge support from STFC consolidated grant numbers ST/R000824/1 and ST/V000861/1, and UKSA grant number ST/R003203/1.
    Some of the data presented in this paper were obtained from the Mikulski Archive for Space Telescopes (MAST). STScI is operated by the Association of Universities for Research in Astronomy, Inc., under NASA contract NAS5-26555. Support for MAST for non-HST data is provided by the NASA Office of Space Science via grant NNX13AC07G and by other grants and contracts. This paper includes data collected by the TESS mission. Funding for the TESS mission is provided by the NASA's Science Mission Directorate. We acknowledge the use of public TESS data from pipelines at the TESS Science Office and at the TESS Science Processing Operations Center. Resources supporting this work were provided by the NASA High-End Computing (HEC) Programme through the NASA Advanced Supercomputing (NAS) Division at Ames Research Center for the production of the SPOC data products.
\end{acknowledgements}

\bibliography{literature}

\appendix

\section{Additional figure}

\noindent\begin{minipage}{\textwidth}
    \centering
    \includegraphics[width=0.55\textwidth]{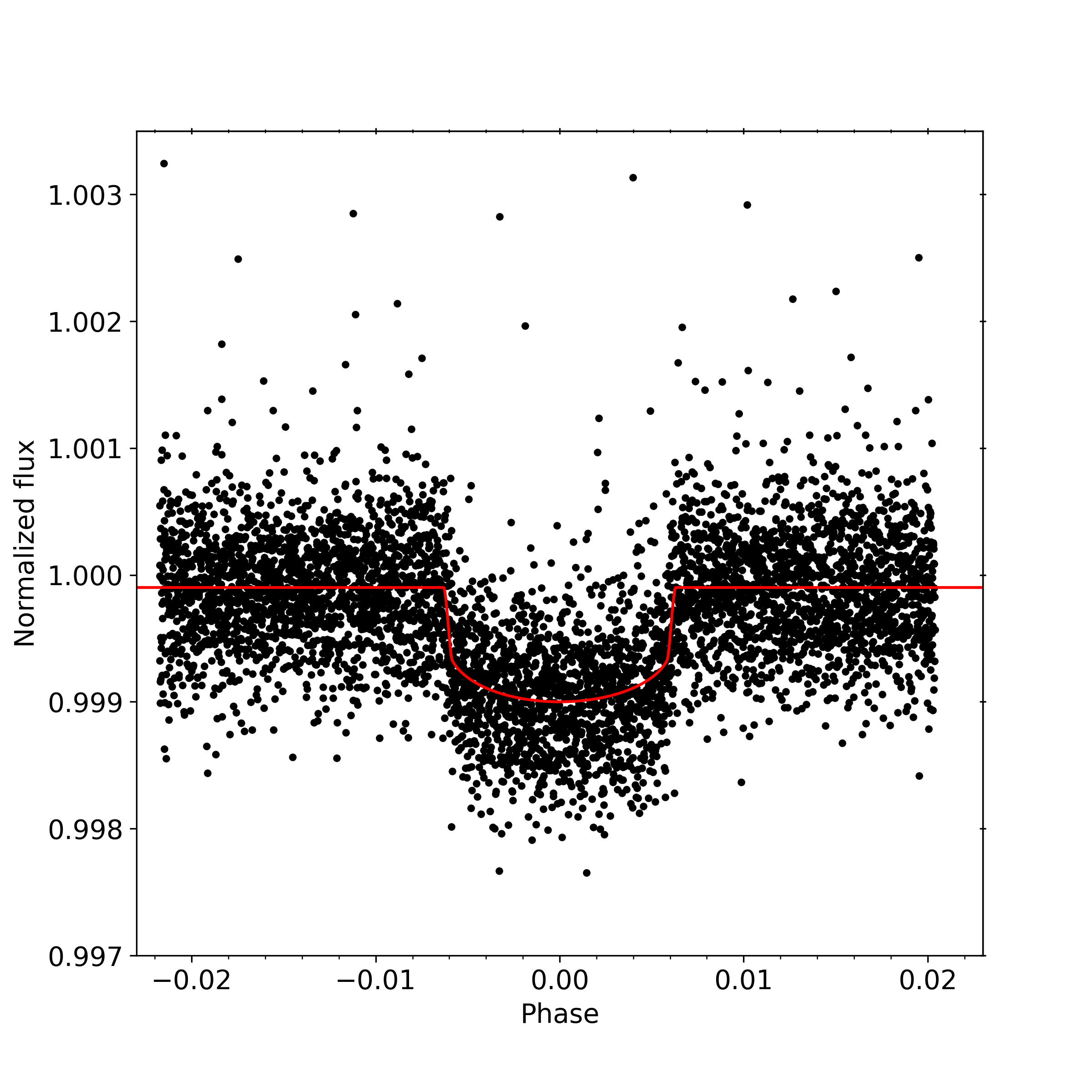}
    \captionof{figure}{HD\,97658\,b phase-folded TESS transit data (black dots) that were previously unpublished. The red line corresponds to the model resulting from the fits to the three available transits in the SAP flux with \textsc{tlcm}.\label{fig:HD97658_tess_transits}}
\end{minipage}
\clearpage
\section{Additional table}
\noindent\begin{minipage}{\textwidth}
    \centering
    \setlength{\tabcolsep}{10pt} 
    \renewcommand{\arraystretch}{1.05} 
    \captionof{table}{HD\,97658\,b mid-transit times and errors. \label{tab:HD97658_tts}}
    \begin{tabular}{c c c c}
        \hline
        Mid-transit time & Error & Epoch & Source \\
        $[\mathrm{BJD}_\mathrm{TDB}]$ & [d] & & \\\hline
        6361.80500 & 0.00330 & 0  & MOST \citep{2020AJ....159..239G} \\
        6523.12544 & 0.00061 & 17 & Spitzer \citep{2014ApJ...786....2V} \\
        6646.48290 & 0.00110 & 30 & HST/WFC3 \citep{2020AJ....159..239G} \\
        6665.46210 & 0.00120 & 32 & HST/WFC3 \citep{2020AJ....159..239G} \\
        6712.90960 & 0.00240 & 37 & MOST \citep{2020AJ....159..239G} \\
        6864.73938 & 0.00027 & 53 & Spitzer \citep{2020AJ....159..239G} \\
        6883.71892 & 0.00027 & 55 & Spitzer \citep{2020AJ....159..239G} \\
        6893.20742 & 0.00026 & 56 & Spitzer \citep{2020AJ....159..239G} \\
        7082.99324 & 0.00032 & 76 & Spitzer \citep{2020AJ....159..239G} \\
        7092.48243 & 0.00034 & 77 & Spitzer \citep{2020AJ....159..239G} \\
        7101.97164 & 0.00034 & 78 & Spitzer \citep{2020AJ....159..239G} \\
        7234.82179 & 0.00026 & 92 & Spitzer \citep{2020AJ....159..239G} \\
        7244.31157 & 0.00022 & 93 & Spitzer \citep{2020AJ....159..239G} \\
        7253.80001 & 0.00045 & 94 & Spitzer \citep{2020AJ....159..239G} \\
        7472.05445 & 0.00025 & 117 & Spitzer \citep{2020AJ....159..239G} \\
        7481.54364 & 0.00031 & 118 & Spitzer \citep{2020AJ....159..239G} \\
        7491.03120 & 0.00110 & 119 & HST/WFC3 \citep{2020AJ....159..239G} \\
        7785.20210 & 0.00110 & 150 & HST/WFC3 \citep{2020AJ....159..239G} \\
        8904.94070 & 0.00100 & 268 & TESS \citep{2021MNRAS.tmp.3057M} \\
        8961.87669 & 0.00028 & 274 & CHEOPS \\
        9294.00198 & 0.00047 & 309 & CHEOPS \\
        9303.49113 & 0.00041 & 310 & CHEOPS \\
        9607.14869 & 0.00051 & 342 & CHEOPS \\
        9654.59570 & 0.00084 & 347 & TESS \\
        \hline
    \end{tabular}
    \tablefoot{The respective epochs are given relative to the beginning of the data set. The ``Source'' column describes the source of the respective light curve or timing. This is described in more detail in Section~\ref{sec:obs}. The timings without references were obtained by us. This table will also be available with the online version of this paper and at the CDS.}
\end{minipage}

\end{document}